\documentclass{article}%
\usepackage{amsmath}
\usepackage{amsfonts}
\usepackage{amssymb}
\usepackage{graphicx}
\usepackage{subfigure}
\usepackage{epsfig}%
\usepackage{color}
\newtheorem{theorem}{Theorem}

\newtheorem{corollary}[theorem]{Corollary}

\newtheorem{lemma}[theorem]{Lemma}

\newtheorem{proposition}[theorem]{Proposition}

\newenvironment{proof}[1][Proof]{\noindent\textbf{#1.} }{\ \rule{0.5em}{0.5em}}
\begin{document}

\title{Prior-Based Model Checking}
\author{Luai Al-Labadi and Michael Evans\\Department of Statistical Sciences\\University of Toronto}

\date{}
\maketitle

\begin{abstract}
Model checking procedures are considered based on the use of the Dirichlet
process and relative belief. This combination is seen to lead to some unique
advantages for this problem. In particular, it avoids double use of the data
and prior-data conflict. Several examples have been incorporated, in which the
proposed approach exhibits excellent performance.

\end{abstract}

\section{Introduction}

This paper is concerned with checking whether or not a chosen statistical
model $\{P_{\theta}:\theta\in\Theta\}$ is in agreement with observed data
$x\in\mathfrak{X},$ where $\mathfrak{X}$ is the sample space with $\sigma
$-algebra $\mathcal{A}$ and each $P_{\theta}$ is a probability measure on
$\mathcal{A}.$ If it is determined that the observed data does not contradict
the model, then inferences can proceed about the true value of $\theta
\in\Theta$. If the model fails to pass its checks, then there is a concern
about the correctness of the inferences. Thus, checking a proposed model based
on the observed data is a matter of some significance.

While there have been many methods developed for model checking, the approach
taken here is Bayesian in nature in that a prior is placed on the set of all
probability measures on $(\mathfrak{X},\mathcal{A)}$ and inference is then
conducted concerning model correctness. The approach taken to inference is
based on a particular measure of evidence known as the relative belief ratio
which measures how beliefs have changed from a priori to a posteriori. So a
relative belief ratio is computed which indicates whether there is evidence
for or against the model $\{P_{\theta}:\theta\in\Theta\}$ holding.
Furthermore, a calibration of this evidence is provided concerning whether
there is strong or weak evidence for or against the model. Relative belief
ratios and the associated inferences are discussed in Section 2.

Recently, there has been considerable interest in developing Bayesian
nonparametric procedures for model checking. Most of this has focused on
embedding the proposed model as a null hypothesis in a larger family of
distributions. Then priors are placed on the null and the alternative and a
Bayes factor is computed. For example, Florens, Richard, and Rolin (1996) used a Dirichlet process for the prior on
the alternative.  Carota and Parmigiani (1996), Verdinelli and Wasserman (1998), Berger and Guglielmi (2001)
and McVinish, Rousseau, and Mengersen (2009) considered a mixture of Dirichlet processes, a mixture of Gaussian
processes, a mixture of P\'{o}lya trees and a mixture of triangular
distributions, respectively, for the prior on the alternative. Another
approach for model testing is based on placing a prior on the true
distribution generating the data and measuring the distance between the
posterior distribution and the proposed one. Swartz (1999) and Al-Labadi and
Zarepour (2013, 2014) considered the Dirichlet process prior and used the
Kolmogorov distance to derive a goodness-of-fit test for continuous models.
Viele (2000) used the Dirichlet process and the Kullback-Leibler distance to
test only discrete models. Hsieh (2011) used the P\'{o}lya tree prior and the
Kullback-Leibler distance to test continuous distributions.

The methodology developed in this paper combines the previous two approaches
and provides some unique, beneficial features. A Dirichlet process $DP(a,H)$
is considered as a prior on the set of all distributions on $(\mathfrak{X}%
,\mathcal{A)}$ and then the concentration of the posterior distribution about
the model of interest is compared to the concentration of the prior
distribution about the model of interest. This comparison is made via a
relative belief ratio to measure the evidence in the observed data for or
against the model. A measure of the strength of this evidence is also
provided. Implementing the approach is fairly simple and does not require
obtaining a closed form of the relative belief ratio. The methodology does not
require the use of a prior on $\theta$ and so is truly a check on the model
itself avoiding any issues with the prior on $\theta.$ It is shown that, by
appropriate choices of the hyperparameters $a$ and $H,$ prior-data conflict
with respect to $DP(a,H),$ namely, the distributions in the model lie in the
tails of the prior, can be avoided. Any prior on $\theta$ should be checked
for prior-data conflict separately from a check on the model, and only when
the model passes its checks, as this avoids confounding model error with error
introduced by a poor choice of a prior, see Evans and Moshonov (2006).

In Section 3 the Dirichlet process prior $DP(a,H)$ is briefly reviewed and in
Section 4 the basis of our goodness-of-fit measure, namely, the Cram\'{e}r-von
Mises distance between probability measures is discussed. Section 5 deals with
the heart of our proposal where it is argued that a particular usage of the
Cram\'{e}r-von Mises distance together with particular choices of the
hyperparameters $(a,H)$ be employed. In Section 6 a computational algorithm is
developed for the implementation of relative belief inferences in this
context. Section 7 presents a number of examples where the behavior of the
methodology is examined in some detail.

\section{Relative Belief Ratios}

Let $\{f_{\theta}:\theta\in\Theta\}$ denote a collection of densities on a
sample space $\mathcal{X}$ and let $\pi$ denote a prior on $\Theta.$ After
observing data $x,$ the posterior distribution of $\theta$ is given by the
density $\pi(\theta\,|\,x)=\pi(\theta)f_{\theta}(x)/m(x)$ where $m(x)=\int
_{\Theta}\pi(\theta)f_{\theta}(x)\,d\theta$ is the prior predictive density of
$x.$ For an arbitrary parameter of interest $\psi=\Psi(\theta),$ denote the
prior and posterior densities of $\psi$ by $\pi_{\Psi}$ and $\pi_{\Psi}%
(\cdot\,|\,x),$ respectively. The relative belief ratio for a value $\psi$ is
then defined by $RB_{\Psi}(\psi\,|\,x)=\lim_{\delta\rightarrow0}\Pi_{\Psi
}(N_{\delta}(\psi\,)|\,x)/\Pi_{\Psi}(N_{\delta}(\psi\,)$ where $N_{\delta
}(\psi\,)$ is a sequence of neighborhoods of $\psi$ converging (nicely) to
$\psi$ as $\delta\rightarrow0.$ Quite generally
\begin{equation}
RB_{\Psi}(\psi\,|\,x)=\pi_{\Psi}(\psi\,|\,x)/\pi_{\Psi}(\psi), \label{relbel}%
\end{equation}
the ratio of the posterior density to the prior density at $\psi.$ So
$RB_{\Psi}(\psi\,|\,x)$ is measuring how beliefs have changed concerning
$\psi$ being the true value from \textit{a priori} to \textit{a posteriori }by
comparing a posterior probability to a prior probability. Note that a relative
belief ratio is similar to a Bayes factor, as both are measures of evidence,
but the latter measures this via the change in an odds ratio. The full
relationship between relative belief ratios and Bayes factors is discussed in
Evans (2015). Our developments here are based on the relative belief ratio as
the associated theory is much simpler.

By a basic principle of evidence, when $RB_{\Psi}(\psi\,|\,x)>1$ the data have
lead to an increase in the probability that $\psi$ is correct, and so there is
evidence in favor of $\psi,$ when $RB_{\Psi}(\psi\,|\,x)<1$ the data have lead
to a decrease in the probability that $\psi$ is correct, and so there is
evidence against $\psi,$ and when $RB_{\Psi}(\psi\,|\,x)=1$ there is no
evidence either way. Note that $RB_{\Psi}(\psi\,|\,x)$ is invariant under
smooth changes of variable and also invariant to the choice of the support
measure for the densities. As such all relative belief inferences possess this
invariance which is not the case for many Bayesian inferences such as using a
posterior mode or expectation for estimation.

The value $RB_{\Psi}(\psi_{0}\,|\,x)$ then measures the evidence for the
hypothesis $\mathcal{H}_{0}=\{\theta:\Psi(\theta)=\psi_{0}\}.$ It is also
necessary, however, to calibrate whether this is strong or weak evidence for
or against $\mathcal{H}_{0}.$ Certainly the bigger $RB_{\Psi}(\psi_{0}\,|\,x)$
is than 1, the more evidence there is in favor of $\psi_{0}$ while the smaller
$RB_{\Psi}(\psi_{0}\,|\,x)$ is than 1, the more evidence there is against
$\psi_{0}.$ But what exactly does a value of $RB_{\Psi}(\psi_{0}\,|\,x)=20$
mean? It would appear to be strong evidence in favor of $\psi_{0}$ because
beliefs have increased by a factor of 20 after seeing the data. But what if
other values of $\psi$ had even larger increases? A\ useful calibration of
$RB_{\Psi}(\psi_{0}\,|\,x)$ is given by%
\begin{equation}
\Pi_{\Psi}(RB_{\Psi}(\psi\,|\,x)\leq RB_{\Psi}(\psi_{0}\,|\,x)\,|\,x),
\label{strength}%
\end{equation}
namely, the posterior probability that the true value of $\psi$ has a relative
belief ratio no greater than that of the hypothesized value $\psi_{0}.$ Note
that (\ref{strength}) is not a p-value as it has a very different
interpretation. When $RB_{\Psi}(\psi_{0}\,|\,x)<1,$ so there is evidence
against $\psi_{0},$ then a small value for (\ref{strength}) indicates a large
posterior probability that the true value has a relative belief ratio greater
than $RB_{\Psi}(\psi_{0}\,|\,x)$ and so there is strong evidence against
$\psi_{0}.$ When $RB_{\Psi}(\psi_{0}\,|\,x)>1,$ so there is evidence in favor
of $\psi_{0},$ then a large value for (\ref{strength}) indicates a small
posterior probability that the true value has a relative belief ratio greater
than $RB_{\Psi}(\psi_{0}\,|\,x))$ and so there is strong evidence in favor of
$\psi_{0},$ while a small value of (\ref{strength}) only indicates weak
evidence in favor of $\psi_{0}.$

As $RB_{\Psi}(\psi\,|\,x)$ measures the evidence that $\psi$ is the true
value, it naturally leads to an estimate of $\psi.$ For example, the best
estimate of $\psi$ is clearly the value for which the evidence is greatest,
namely, $\psi(x)=\arg\sup RB_{\Psi}(\psi\,|\,x).$ Associated with this is a
$\gamma$-credible region $C_{\Psi,\gamma}(x)=\{\psi:RB_{\Psi}(\psi\,|\,x)\geq
c_{\Psi,\gamma}(x)\}$ where $c_{\Psi,\gamma}(x)=\inf\{k:\Pi_{\Psi}(RB_{\Psi
}(\psi\,|\,x)>k\,|\,x)\leq\gamma\}.$ Notice that $\psi(x)\in C_{\Psi,\gamma
}(x)$ for every $\gamma\in\lbrack0,1]$ and so, for selected $\gamma,$ we can
take the "size" of $C_{\Psi,\gamma}(x)$ as a measure of the accuracy of the
estimate $\psi(x).$ The interpretation of $RB_{\Psi}(\psi\,|\,x)$ as the
evidence for $\psi\ $forces the sets $C_{\Psi,\gamma}(x)$ to be the credible
regions. For if $\psi_{1}$ is in such a region and $RB_{\Psi}(\psi
_{2}\,|\,x)\geq RB_{\Psi}(\psi_{1}\,|\,x),$ then $\psi_{2}$ must also be in
the region as there is at least as much evidence for $\psi_{2}$ as for
$\psi_{1}.$

A number of optimality results have been established for relative belief
inferences and these are discussed in Evans (2015). For example, suppose we
use the relative belief ratio to accept $\mathcal{H}_{0}:\Psi(\theta)=\psi
_{0}$ when $RB_{\Psi}(\psi_{0}\,|\,x)>1$ and reject when $RB_{\Psi}(\psi
_{0}\,|\,x)<1.$ It is the case then that the acceptance region $A(\psi
_{0})=\{x:RB_{\Psi}(\psi_{0}\,|\,x)>1\}$ and the rejection region $R(\psi
_{0})=\{x:RB_{\Psi}(\psi_{0}\,|\,x)<1\}$ are optimal among all such regions in
the following sense. Let $A\subset\mathcal{X}$ be another acceptance region
such that $M(A\,|\,\psi_{0})\geq M(A(\psi_{0})\,|\,\psi_{0})$ where
$M(\cdot\,|\,\psi_{0})$ is the conditional prior predictive probability
measure given that $\Psi(\theta)=\psi_{0}.$ Then among all such acceptance
regions, $A(\psi_{0})$ minimizes the prior probability of rejecting $H_{0}$
when it is false. A similar result holds for $R(\psi_{0}).$ Furthermore, under
mild conditions it is proved in Evans (2015) that $M(A(\psi_{0})\,|\,\psi
_{0})\rightarrow1$ and $M(R(\psi_{0})\,|\,\psi_{0})\rightarrow0$ as the amount
of data increases. So the values of $M(A(\psi_{0})\,|\,\psi_{0})$ and
$M(R(\psi_{0})\,|\,\psi_{0})$ can be set by design and it is then known that
we are using the optimal tests with these characteristics. Numerous additional
optimality results are proved for the relative credible regions $C_{\Psi
,\gamma}(x)$ and the estimator $\psi(x)$ in Evans (2015).

The view is taken here that anytime continuous probability is used, then this
is an approximation to a finite, discrete context. For example, if $\psi$ is a
mean and the response measurements are to the nearest centimeter, then of
course the true value of $\psi$ cannot be known to an accuracy greater than
1/2 of a centimeter, no matter how large a sample we take. Furthermore, there
are implicit bounds associated with any measurement process. As such the
restriction is made here to discretized parameters that take only finitely
many values. So when $\psi$ is a continuous, real-valued parameter, it is
discretized to the intervals $\ldots,(\psi_{0}-3\delta,\psi_{0}-\delta
],(\psi_{0}-\delta,\psi_{0}+\delta],(\psi_{0}+\delta,\psi_{0}+3\delta],\ldots$
for some choice of $\delta>0,$ and there are only finitely may such intervals
covering the range of possible values. It is of course possible to allow the
intervals to vary in length as well. With this discretization, then $H_{0}=$
$(\psi_{0}-\delta,\psi_{0}+\delta].$

Note that throughout the paper the notation $P$ could refer to either a
probability measure or its corresponding cdf where the context determines the
appropriate interpretation.

\section{Dirichlet Process}

The Dirichlet process, formally introduced in Ferguson (1973), is the most
well-known and widely used prior in Bayesian nonparametric inference. Consider
a space $\mathfrak{X}$ with a $\sigma-$algebra $\mathcal{A}$ of subsets of
$\mathfrak{X}$. Let $H$ be a fixed probability measure on $(\mathfrak{X}%
,\mathcal{A})$ and $a$ be a positive number. Following Ferguson (1973), a
random probability measure $P=\left\{  P(A)\right\}  _{A\in\mathcal{A}}$ is
called a Dirichlet process on $(\mathfrak{X},\mathcal{A})$ with parameters $a$
and $H$, if for any finite measurable partition $\{A_{1},\ldots,A_{k}\}$ of
$\mathfrak{X}$, the joint distribution of the vector $\left(  P(A_{1}%
),\ldots\,P(A_{k})\right)  $ has the Dirichlet distribution with parameters
$(aH(A_{1}),\ldots,$ $aH(A_{k})),$ where $k\geq2$. We assume that if
$H(A_{j})=0$, then $P(A_{j})=0$ with a probability one. If $P$ is a Dirichlet
process with parameters $a$ and $H,$ we write $P\sim{DP}(a,H).$ For any
$A\in\mathcal{A},$ $P(A)$ has a beta distribution with parameters $aH(A)$ and
$a(1-H(A))$ and so ${E}(P(A))=H(A)\ \ $ and ${Var}(P(A))=H(A)(1-H(A))/(1+a).$
The probability measure $H$ is called the \emph{base measure} of $P$. Clearly
$H$ plays the role of the \emph{center} of the process, while $a$ can be
viewed as the \emph{concentration parameter}. The larger $a$ is, the more
likely it is that the realization of $P$ is close to $H$.

An attractive feature of the Dirichlet process is the conjugacy property. If
$x=(x_{1},\ldots,x_{n})$ is a sample from $P\sim DP(a,H)$, then the posterior
distribution of $P$ is $P\,|\,x\sim DP(a+n,H_{x})$ where
\begin{equation}
H_{x}=a(a+n)^{-1}H+n(a+n)^{-1}F_{n}, \label{DP_posterior}%
\end{equation}
with $F_{n}=n^{-1}\sum_{i=1}^{n}\delta_{{x}_{i}}$ and $\delta_{x_{i}}$ the
Dirac measure at $x_{i}.$ Notice that, the posterior base distribution $H_{x}$
is a convex combination of the prior base distribution and the empirical
distribution. The posterior base $H_{x}$ approaches the prior base $H$ as
$a\rightarrow\infty$ while $H_{x}$ converges to the empirical distribution as
$a\rightarrow0.$

Ferguson (1973) provided a series representation for $P\sim{DP}(a,H).$
Specifically, let $(E_{k})_{k\geq1}$ be i.i.d. exponential$(1)$ random
variables, $\Gamma_{i}=E_{1}+\cdots+E_{i},(Y_{i})_{i\geq1}$ be i.i.d. $H$
random variables independent of $(\Gamma_{i})_{i\geq1}$ and put
\begin{equation}
P=\sum_{i=1}^{\infty}L^{-1}(\Gamma_{i}){\delta_{Y_{i}}/}\sum_{i=1}^{\infty
}{{L^{-1}(\Gamma_{i})}}, \label{series-dp}%
\end{equation}
where $L(x)=a\int_{x}^{\infty}t^{-1}e^{-t}dt,x>0,$ and $L^{-1}(y)=\inf
\{x>0:L(x)\geq y\}.$ From (\ref{series-dp}), it follows clearly that a
realization of the Dirichlet process is a discrete probability measure. This
is true even when the base measure is absolutely continuous. Note that,
although the Dirichlet process is discrete with probability one, this
discreteness is no more troublesome than the discreteness of the empirical
process. By imposing the weak topology, the support for the Dirichlet process
is quite large. Specifically, the support for the Dirichlet process is the set
of all probability measures whose support is contained in the support of the
base measure. This means if the support of the base measure is $\mathfrak{X}$,
then the space of all probability measures is the support of the Dirichlet
process. For example, if we have a normal base measure, then the Dirichlet
process can choose any probability measure.

Recently, Zarepour and Al-Labadi (2012) derived an efficient series
approximation with monotonically decreasing weights for the Dirichlet process
. Let $(Y_{i})_{1\leq i\leq N}$ be i.i.d. $H$ independent of $(\Gamma
_{i})_{1\leq i\leq{N+1}},G_{a/N}$ be the co-cdf of the g$\text{amma}(a/N,1)$
distribution, and $J_{i}={G_{a/N}^{-1}(\Gamma_{i}/\Gamma_{N+1})/}\sum
_{j=1}^{N}{G_{a/N}^{-1}(\Gamma_{j}/\Gamma_{N+1}),}$ then%
\begin{equation}
P_{N}=\sum_{i=1}^{N}J_{i}\delta_{Y_{i}} \label{eq11}%
\end{equation}
converges almost surely to $P$ defined by (\ref{series-dp}), as $N\rightarrow
\infty$. Note that ${G_{a/N}^{-1}(p)}$ is the $(1-p)$-th quantile of the
g$\text{amma}(\alpha/N,1)$ distribution. This provides the following
algorithm.$\bigskip$

\noindent\textbf{Algorithm A: Approximately generating a value from }%
$DP(a,H)$\textbf{\smallskip}

\noindent1. Fix a relatively large positive integer $N$.\textbf{\smallskip}

\noindent2. Generate i.i.d. $Y_{i}\sim H$ for $i=1,\ldots,N.$%
\textbf{\smallskip}

\noindent3. For $i=1,\ldots,N+1,$ generate i.i.d. $E_{i}\sim\,$%
exponential$(1)$ distribution independent of $\left(  Y_{i}\right)  _{1\leq
i\leq N}$ and put $\Gamma_{i}=E_{1}+\cdots+E_{i}.$\textbf{\smallskip}

\noindent4. For $i=1,\ldots,N$ compute $G_{a/N}^{-1}\left(  {\Gamma_{i}%
}/{\Gamma_{N+1}}\right)  .$\textbf{\smallskip}

\noindent6. Use (\ref{eq11}) to obtain an approximate value from
$DP(a,H)$.$\bigskip$

\noindent For other simulation methods for the Dirichlet process, see
Bondesson (1982), Sethuraman (1994), and Wolpert and Ickstadt (1998).

\section{Cram\'{e}r-von Mises Distance}

A widely used distance between distributions is the Cram\'{e}r-von Mises
distance. For cdf's $F$ and $G$ this is defined as $d_{CvM}(F,G)=\int
_{-\infty}^{\infty}\left(  F(x)-G(x)\right)  ^{2}G(dx).$ Note that other
distances could be employed in our analysis, see Gibbs and Su (2002), but
$d_{CvM}$ has some convenient attributes.

The following lemma, as given in Al-Labadi and Zarepour (2014), provides a
simple formula for the distance between a discrete and a continuous cdf.

\begin{lemma}
\label{cvm1} Let $G$ be a continuous cdf and $P_{N}=\sum_{i=1}^{N}J_{i}%
\delta_{Y_{i}}$ be a discrete distribution, where $Y_{(1)}\leq\ldots\leq
Y_{(N)}$ are the order statistics of $(Y_{i})_{1\leq i\leq N}$ and
$J_{1}^{\prime},\dots,J_{N}^{\prime}$ are the associated jump sizes such that
$J_{i}=J_{j}^{\prime}$ when $Y_{i}=Y_{(j)}.$ Then
\[
d_{CvM}(P_{N},G)=1/3+\sum_{i=1}^{N}J_{i}^{\prime}G^{2}(Y_{(i)})-\sum_{i=1}%
^{N}{J^{\prime}}_{i}^{2}G(Y_{(i)})-2\sum_{i=2}^{N}J_{i}^{\prime}\sum
_{k=1}^{i-1}J_{k}^{\prime}G(Y_{(i)}).
\]

\end{lemma}

\noindent A corollary gives that the distribution of $d_{CvM}(P_{N},G)$ is
independent of $G$ whenever $H=G$ and $P_{N}=\sum_{i=1}^{N}J_{i}\delta_{Y_{i}%
}$.

\begin{corollary}
\label{cvm2}Suppose that $(Y_{i})_{1\leq i\leq N}\sim G$ are i.i.d.,
independent of $(J_{i})_{1\leq i\leq N}$ and $P_{N}=\sum_{i=1}^{N}J_{i}%
\delta_{Y_{i}}$. Then $d_{CvM}(P_{N},G)\overset{d}{=}1/3+\sum_{i=1}^{N}%
J_{i}^{\prime}U_{(i)}^{2}-\sum_{i=1}^{N}{J^{\prime}}_{i}^{2}U_{(i)}%
-2\sum_{i=2}^{N}J_{i}^{\prime}\sum_{k=1}^{i-1}J_{k}^{\prime}U_{(i)}$ where
$U_{(i)}$ is the $i$-th order statistic for $\left(  U_{i}\right)  _{1\leq
i\leq N}$ i.i.d. uniform$[0,1].$
\end{corollary}

\proof Since $(Y_{i})_{1\leq i\leq N}$ is a sequence of i.i.d. random
variables with continuous distribution $G$, then $(U_{i})_{1\leq i\leq
N}=(G(Y_{i}))_{1\leq i\leq N}$ is i.i.d. uniform$[0,1]$ and the result follows
from Lemma \ref{cvm1}. \endproof

The following result allows the use of the approximation to the Dirichlet
process when considering the prior and posterior distributions of the
Cram\'{e}r-von Mises distance.

\begin{lemma}
\label{cvm3}If $P\sim DP(a,H)$ and $P_{N}$ is given by (\ref{eq11}), then
$d_{CvM}\left(  P_{N},G\right)  \overset{a.s.}{\rightarrow}d_{CvM}\left(
P,G\right)  $ as $N\rightarrow\infty.$
\end{lemma}

\proof This follows by the dominating convergence theorem since for any cdf's
$G$ and $H,(P_{N}(x)-G(x))^{2}\leq1,d_{CvM}\left(  P_{N},G\right)
=\int_{-\infty}^{\infty}\left(  P_{N}(x)-G(x)\right)  ^{2}G(dx)$ and
$P_{N}(x)\overset{a.s.}{\rightarrow}P(x)$.

\section{Relative Belief Approach for Model Checking}

Let $\left\{  F_{\theta}:\theta\in\Theta\right\}  $ denote the collection of
cumulative distribution functions for the model and assume hereafter that
these are continuous. Suppose that $x=(x_{1},\ldots,x_{n})$ is a sample from a
distribution $P$ and the aim is to test the hypothesis $\mathcal{H}_{0}%
:P\in\left\{  F_{\theta}:\theta\in\Theta\right\}  $. To this end, we use the
prior $P\sim DP(a,H)$ for some choice of $a$ and $H$ so, by
(\ref{DP_posterior}), $P\,|\,x\sim DP\left(  a+n,H_{x}\right)  $. If
$\mathcal{H}_{0}$ is true, then we expect the observed data to lead to the
posterior distribution of the distance between $P$ and $\left\{  F_{\theta
}:\theta\in\Theta\right\}  $ being more concentrated about $0$ than the prior
distribution of the distance between $P$ and $\left\{  F_{\theta}:\theta
\in\Theta\right\}  .$ For example, Figure 1-a (see Example 1) is a plot of the
prior and posterior densities of $d_{CvM}$ in a case where $\mathcal{H}_{0}$
is true and indeed the posterior is much more concentrated about 0 than the
prior. So our test will involve a comparison of the concentrations of the
prior and posterior distributions of $d_{CvM}$ via a relative belief ratio
based on $d_{CvM}$ with the interpretation as discussed in$\ $Section 2.

The first step is to determine how $d_{CvM}$ is to be used to measure the
concentration of the prior and posterior about $\left\{  F_{\theta}:\theta
\in\Theta\right\}  .$ One possibility is to look at the prior and posterior
distributions of $\inf\{d_{CvM}(P,F_{\theta}):\theta\in\Theta\}.$ While this
is reasonable, a simpler approach, that avoids the computation of the infimum,
is to choose the distribution $F_{\theta}$ which is best supported by the data
and look at the prior and posterior distributions of $d_{CvM}(P,F_{\theta})$
as a measure of the closeness of $P$ to $\left\{  F_{\theta}:\theta\in
\Theta\right\}  .$ Of course, when using relative belief ratios to measure
evidence, the $F_{\theta}$ that is best supported by the data is
$F_{\theta(x)},$ where $\theta(x)$ is the relative belief estimate of $\theta
$. Note that the relative belief estimate of the full parameter $\theta$ is
also the MLE and this is independent of any prior $\Pi$ on $\theta.$ This
would appear to induce a data dependent prior distribution for $d_{CvM}$ but
in fact this is not the case for the approach developed here. This is
accomplished by letting $H=F_{\theta(x)}$ in the $DP(a,H)$ prior so the lack
of dependence on the data is immediate from Corollary \ref{cvm2}. So,
considering the space of all probability measures $P$ on $(\mathfrak{X}%
,\mathcal{A}),$ we take $d=D(P)=d_{CvM}(P,F_{\theta(x)})$ and assess
$\mathcal{H}_{0}$ using $RB_{D}(0\,|\,x)$ and its corresponding strength.
Lemma \ref{cvm4} justifies this approach. From Lemma \ref{cvm3}, note that the
prior distribution of $d_{CvM}(P,F_{\theta(x)})$ can be approximated by the
prior distribution of $d_{CvM}(P_{N},F_{\theta(x)}).$

There is another reason why choosing $H=F_{\theta(x)}$ makes sense. For,
whatever choice of $H$ is made, it is necessary to avoid prior-data conflict
as discussed, for example, in Evans and Moshonov (2006). Prior-data conflict
here means that every $F_{\theta}$ lies in the "tails" of $DP(a,H).$ While it
is true that the effect of the prior is overwhelmed by large amounts of data,
for small sample sizes the prior can seriously distort things. In this
context, when prior-data conflict exists, there can fail to be an appreciable
concentration of the posterior distribution of $d_{CvM}\left(  P,F_{\theta
(x)}\right)  $ about 0 even when $\mathcal{H}_{0}$ is true. Prior-data
conflict will occur whenever there is a only tiny overlap between the
effective support regions of $P$ and $F_{\theta(x)}$. Specifically, by Lemma
\ref{cvm1}, $d_{CvM}\left(  P,F_{\theta(x)}\right)  $ depends on the base
measure $H$ through the jump points $Y_{i}$. If the $Y_{i}$ lie in one tail of
$F_{\theta(x)}$, then we get prior-data conflict between $P$ and
$F_{\theta(x)}$ as $H$ and $P$ have the same effective support. To avoid this
it is necessary that the $Y_{i}$ are selected in a region that contains most
of the mass of $F_{\theta(x)}.$ Note that when $H=F_{\theta(x)}$ then
$F_{\theta(x)}$ is the prior mean of $P$ and thus both share the same
effective support. The effect of prior-data conflict is demonstrated in
Example 1.

The choice of $H$ should also avoid any effects due to "double use of the
data". Such an effect typically means that the methodology results in overly
conservative outcomes such that model failure is not detected when
$\mathcal{H}_{0}$ is false. To see that this is not the case when
$H=F_{\theta(x)},$ it is now established that the posterior distribution of
$d_{CvM}(P,F_{\theta(x)})$ becomes concentrated around 0 as sample size
increases if and only if $\mathcal{H}_{0}$ holds. Throughout the remainder of
this paper $\theta_{0}$ is the value that minimizes the divergence between the
true distribution and a member of $\left\{  F_{\theta}:\theta\in
\Theta\right\}  .$

\begin{lemma}
\label{cvm4}Let $P\sim DP(a+n,H_{x})$ and suppose that $\theta(x)\overset
{a.s.}{\rightarrow}\theta_{0},\sup_{z}|F_{\theta(x)}(z)-F_{\theta_{0}%
}(z)|\overset{a.s.}{\rightarrow}0$ as $n\rightarrow\infty.$ (i) If
$\mathcal{H}_{0}$ is true, then $d_{CvM}\left(  P,F_{\theta(x)}\right)
\overset{a.s.}{\rightarrow}0$ and (ii) if $\mathcal{H}_{0}$\ is false, then
$\lim\inf d_{CvM}(P,F_{\theta(x)})\overset{a.s.}{>}0.$
\end{lemma}

\proof(i) Since $d_{CvM}(F,G)\leq\sup_{z\in\mathbb{R}}|F(z)-G(z)|$, then the
triangle inequality implies $d_{CvM}\left(  P,F_{\theta(x)}\right)  \leq
\sup_{z\in\mathbb{R}}|P(z)-H_{x}(z)|+\sup_{z\in\mathbb{R}}|H_{x}%
(z)-F_{\theta(x)}(z)|.$ The result follows, as $\sup_{z\in\mathbb{R}%
}|P(z)-H_{x}(z)|\overset{a.s.}{\rightarrow}0$ as $n\rightarrow\infty$ from
James (2008), and $\sup_{x\in\mathbb{R}}|H_{x}(z)-F_{\theta(x)}(z)|\overset
{a.s.}{\rightarrow}0$ under $H_{0}$ by the continuous mapping theorem and
Poly\'{a}'s theorem, see Dasgupta (2008). (ii) As proved in Choi and Bulgren
(1968), $3d_{CvM}(P,F_{\theta(x)})\geq\left(  \sup_{z\in\mathbb{R}%
}|P(z)-F_{\theta(x)}(z)|\right)  ^{3}.$ Using the triangle inequality,
$\sup_{z\in\mathbb{R}}|H_{x}-F_{\theta(x)}(z)|-\sup_{z\in\mathbb{R}%
}|P(z)-H_{x}(z)|\leq\sup_{z\in\mathbb{R}}|P(z)-F_{\theta(x)}(z)|.$ Again
$\sup_{z\in\mathbb{R}}|P(z)-H_{x}(z)|\overset{a.s.}{\rightarrow}0\ $and, since
$\mathcal{H}_{0}$ doesn't hold, $\sup_{z\in\mathbb{R}}|H_{x}-F_{\theta
(x)}(z)|\overset{a.s.}{\rightarrow}\sup_{z\in\mathbb{R}}|F_{true}%
(z)-F_{\theta_{0}}(z)|>0.$ Therefore, $\lim\inf\sup_{z\in\mathbb{R}%
}|P(z)-F_{\theta(x)}(z)|\overset{a.s.}{>}0$ which implies $\lim\inf
d_{CvM}(P,F_{\theta(x)})\overset{a.s.}{>}0$. \endproof

The hyperparameter $a$ also needs to be chosen and so its effect needs to be
studied. For this let $P_{N}^{\ast}=\sum_{i=1}^{N}J_{i,N}\delta_{{Y}_{i}}$
denote the finite dimensional approximation of the $DP(a,H)$ process developed
in Ishwaran and Zarepour (2002), where $(Y_{i})_{i\geq1}$ is i.i.d. $H$
independent of $(J_{i,N})_{1\leq i\leq N}\sim\,$Dirichlet$(a/N,\ldots,a/N)$.
Then $E_{P_{N}^{\ast}}(g)\rightarrow E_{P}(g)$ in distribution as
$N\rightarrow\infty$, for any measurable function $g:\mathbb{R}\rightarrow
\mathbb{R}$ with $\int_{\mathbb{R}}|g(x)|H(dx)<\infty$ and $P\sim DP(a,H)$. In
particular, $(P_{N}^{\ast})_{N\geq1}$ converges in distribution to $P$, where
$P_{N}^{\ast}$ and $P$ are random values in the space $M_{1}(\mathbb{R})$ of
probability measures on $\mathbb{R}$ endowed with the topology of weak
convergence. To generate $(J_{i,N})_{1\leq i\leq N}$ put $J_{i,N}=G_{i,N}%
/\sum_{i=1}^{N}{G_{i,N},}$ where $(G_{i,N})_{1\leq i\leq N}$ is a sequence of
i.i.d. gamma$(a/N,1)$ random variables independent of $(Y_{i})_{1\leq i\leq
N}$. This leads to the following result.

\begin{lemma}
\label{cvm5}If $P\sim DP(a,F_{\theta(x)})$ and $P_{N}^{\ast}=\sum_{i=1}%
^{N}J_{i,N}\delta_{{Y}_{i}},$ then (i) $E_{P_{N}^{\ast}}(d_{CvM}$%
\newline$\left(  P_{N}^{\ast},F_{\theta(x)}\right)  =(1-a/N)/6(a+1)$ and (ii)
$E_{P}\left(  d_{CvM}\left(  P,F_{\theta(x)}\right)  \right)  =1/6(a+1).$
\end{lemma}

\proof The result in Lemma \ref{cvm1} applied to $P_{N}^{\ast}$ implies
$E_{P_{N}^{\ast}}\left(  d_{CvM}\left(  P_{N}^{\ast},F_{\theta(x)}\right)
\right)  =1/3+\sum_{i=1}^{N}E_{P_{N}^{\ast}}(J_{i,N}^{\prime})E_{P_{N}^{\ast}%
}(F_{\theta(x)}^{2}(Y_{(i)}))-\sum_{i=1}^{N}E_{P_{N}^{\ast}}({J^{\prime}%
}_{i,N}^{2})E_{P_{N}^{\ast}}(F_{\theta(x)}(Y_{(i)}))-2\sum_{i=2}^{N}%
E_{P_{N}^{\ast}}(J_{i,N}^{\prime}\sum_{k=1}^{i-1}J_{k,N}^{\prime}%
)E_{P_{N}^{\ast}}(F_{\theta(x)}(Y_{(i)}))$. Furthermore, from properties of
the Dirichlet, $E\left(  J_{i,N}\right)  =1/N,E\left(  J_{i,N}^{2}\right)
=(a+N)/N^{2}(a+1),E\left(  J_{i,N}J_{j,N}\right)  =a/N^{2}(a+1)$ and, as
$F_{\theta(x)}\left(  Y_{(i)}\right)  \sim\,$beta$(i,N-i+1)$ independent of
$(J_{i,N})_{1\leq i\leq N},$ then $E\left(  d_{CvM}\left(  P_{N}^{\ast
},F_{\theta(x)}\right)  \right)  =1/3+\sum_{i=1}^{N}%
i(i+1)/N(N+1)(N+2)-(\left(  a/N\right)  +1)\sum_{i=1}^{N}i/N(N+1)(a+1)-2a\sum
_{i=2}^{N}i(i-1)/N^{2}(N+1)(a+1).$ The identities $\sum_{i=1}^{N}%
i(i+1)=N(N+1)(N+2)/3,\sum_{i=1}^{N}i=N(N+1)/2$ and $\sum_{i=2}^{N}%
i(i-1)=N(N+1)(N-1)/3$ establish (i). Taking the limit in $E_{P_{N}^{\ast}%
}\left(  d_{CvM}\left(  P_{N}^{\ast},F_{\theta(x)}\right)  \right)  $ as
$N\rightarrow\infty,$ and using Lemma \ref{cvm3} and dominated convergence
gives (ii). \endproof

\noindent Note that, from Lemma \ref{cvm5}(ii), $E\left(  d_{CvM}\left(
P,F_{\theta(x)}\right)  \right)  \rightarrow0$ as $a\rightarrow\infty$.

The selection of $a$ is an important step in determining the success of the
algorithm. This is dependent on an number of criteria. For example, if $F$
corresponds to $t_{3}$ distribution, namely, a $t$ distribution on 3 degrees
of freedom, and $\{F_{\theta}:\theta\in\Theta\}$ is the location-scale normal
family, then $\inf_{\theta\in\Theta}d_{CvM}(F,F_{\theta})=0.0120$ while when
$F$ is $t_{1}$, then $\inf_{\theta\in\Theta}d_{CvM}(F,F_{\theta})=0.0335.$
Clearly then, the methodology discussed here will have more problems detecting
model failure when the true distribution is like a $t_{3}$ than like a
$t_{1}.$ A natural approach then, to selecting a relevant $a,$ is to first
determine what kind of deviations from $\{F_{\theta}:\theta\in\Theta\}$ it is
desired to detect, for example, a $t_{3}$ distribution in the context of
assessing normality, and then run a simulation study to determine what values
of $a$ are needed to detect this. In principal larger values of $a$ must be
chosen to detect smaller deviations. This issue is further discussed in
Section 7.

It is also possible to consider several values of $a$. For example, one may
start with $a=1$. If the relative belief ratio is less than 1, then this is
evidence against $\mathcal{H}_{0}$ and larger values of $a$ will tend to
reinforce this. On the other hand, if the relative belief ratio is greater
than 1, one may also consider larger values of $a$ to see if a more
concentrated prior produces the same evidence. It is recommended that
$a\leq0.25n,$ however, else the prior may become too influential. If, as the
value of $a$ is increased, the corresponding relative belief ratio drops
rapidly below 1, then this is a clear indication against $\mathcal{H}_{0}$. As
will be seen in the examples, when the model is correct, the relative belief
ratio always remains above 1 when larger values of $a$ are considered.

\section{Computations}

Closed forms of the prior and posterior densities of $d=D(P)=d_{CvM}%
(P,F_{\theta(x)})$ are typically not available and these are necessary if
using (\ref{relbel}) to compute $RB_{D}(d\,|\,x)$. As such the relative belief
ratios need to be approximated via simulation. A special problem arises here
as $\mathcal{H}_{0}$ corresponds (approximately) to $d_{CvM}(P,F_{\theta
(x)})=0$ and both $\pi_{D}(0\,|\,x)\approx0$ and $\pi_{D}(0)\approx0,$ see
Figures 1 and 2. In such a case determining $RB_{D}(0\,|\,x)$ precisely is
difficult. The formal definition of $RB_{D}(0\,|\,x),$ however, as given in
Section 2, is as a limit and this limit can be approximated by $RB_{D}%
([0,d_{\ast})\,|\,x)$, the ratio of the posterior to prior probability that
$0\leq D\leq d_{\ast},$ for a suitably small value of $d_{\ast}.$ In general
$d_{\ast}$ can be chosen to be $d_{p_{0}},$ the $p_{0}$-th quantile of the
prior distribution of $D,$where $p_{0}>0$ is chosen close to 0.

The following gives a computational algorithm for the evidence, and its
strength, for $\mathcal{H}_{0}$. Of necessity this requires a discretization
of the range of possible values for $D$ and this is chosen here to be based on
quantiles of the prior distribution of $D.\bigskip$

\noindent\textbf{Algorithm B: Relative belief algorithm for model
checking\smallskip}

\noindent1. Use Algorithm A to (approximately) generate a $P$ from
$DP(a,F_{\theta(x)})$.\textbf{\smallskip}

\noindent2. Compute $d=d_{CvM}(P,F_{\theta(x)})$.\textbf{\smallskip}

\noindent3. Repeat steps (1)-(2) to obtain a sample of $r_{1}$ values from the
prior of $D$.\textbf{\smallskip}

\noindent4. Use Algorithm A to (approximately) generate a $P$ from
$DP(a+n,H_{x})$.\textbf{\smallskip}

\noindent5. Compute $d=d_{CvM}(P,F_{\theta(x)})$.\textbf{\smallskip}

\noindent6. Repeat steps (4)-(5) to obtain a sample of $r_{2}$ values from the
posterior of $D$.\textbf{\smallskip}

\noindent7. Let $M$ be a positive number. Let $\hat{F}_{D}$ denote the
empirical cdf of $D$ based on the prior sample in (3) and for $i=0,\ldots,M,$
let $\hat{d}_{i/M}$ be the estimate of $d_{i/M},$ the $(i/M)$-th prior
quantile of $D.$ Here $\hat{d}_{0}=0$, and $\hat{d}_{1}$ is the largest value
of $d$. Let $\hat{F}_{D}(\cdot\,|\,x)$ denote the empirical cdf of $D$ based
on the posterior sample in (6). For $d\in\lbrack\hat{d}_{i/M},\hat
{d}_{(i+1)/M})$, estimate $RB_{D}(d\,|\,x)$ by
\begin{equation}
\widehat{RB}_{D}(d\,|\,x)=M\{\hat{F}_{D}(\hat{d}_{(i+1)/M}\,|\,x)-\hat{F}%
_{D}(\hat{d}_{i/M}\,|\,x)\}, \label{rbest}%
\end{equation}
the ratio of the estimates of the posterior and prior contents of $[\hat
{d}_{i/M},\hat{d}_{(i+1)/M}).$ Also, estimate $RB_{D}(0\,|\,x)$ by
$\widehat{RB}_{D}(0\,|\,x)=M\widehat{F}_{D}(\hat{d}_{p_{0}}\,|\,x)$ where
$p_{0}=i_{0}/M$ and $i_{0}$ is chosen so that $i_{0}/M$ is not too small
(typically $i_{0}/M\approx0.05)$.\textbf{\smallskip}

\noindent8. Estimate the strength $DP_{D}(RB_{D}(d\,|\,x)\leq RB_{D}%
(0\,|\,x)\,|\,x)$ by the finite sum
\begin{equation}
\sum_{\{i\geq i_{0}:\widehat{RB}_{D}(\hat{d}_{i/M}\,|\,x)\leq\widehat{RB}%
_{D}(0\,|\,x)\}}(\hat{F}_{D}(\hat{d}_{(i+1)/M}\,|\,x)-\hat{F}_{D}(\hat
{d}_{i/M}\,|\,x)). \label{strest}%
\end{equation}

\noindent For fixed $M,$ as $r_{1}\rightarrow\infty,r_{2}\rightarrow\infty,$
then $\hat{d}_{i/M}$ converges almost surely to $d_{i/M}$ and (\ref{rbest})
and (\ref{strest}) converge almost surely to $RB_{D}(d\,|\,x)$ and
$DP_{D}(RB_{D}(d\,|\,x)\leq RB_{D}(0\,|\,x)\,|\,x)$, respectively.

The following establishes the consistency of the approach to testing the model
as sample size increases.

\begin{proposition}
\label{cvm6}Consider the discretization $\{[0,d_{i_{0}/M}),[d_{i_{0}%
/M},d_{(i_{0}+1)/M}),\ldots,$\newline$[d_{(M-1)/M},\infty)\}$. As
$n\rightarrow\infty,$ (i) if $\mathcal{H}_{0}$ is true, then
\begin{align*}
&  RB_{D}([0,d_{i_{0}/M})\,|\,x)\overset{a.s.}{\rightarrow}1/DP_{D}%
([0,d_{i_{0}/M})),\\
&  RB_{D}([d_{i/M},d_{(i+1)/M})\,|\,x)\overset{a.s.}{\rightarrow}0\text{
whenever }i\geq i_{0},\\
&  DP_{D}(RB_{D}(d\,|\,x)\leq RB_{D}(0\,|\,x)\,|\,x)\overset{a.s.}%
{\rightarrow}1,
\end{align*}
and (ii) if $\mathcal{H}_{0}$ is false and $d_{CvM}(P_{true},F_{\theta_{0}%
})\geq d_{i_{0}/M}$, then $RB_{D}([0,d_{i_{0}/M})\,|\,x)\overset
{a.s.}{\rightarrow}0$ and $DP_{D}(RB_{D}(d\,|\,x)\leq RB_{D}%
(0\,|\,x)\,|\,x)\overset{a.s.}{\rightarrow}0.$
\end{proposition}

\proof These results follow immediately from Evans (2015), Section 4.7.1.
\endproof

\noindent So the procedure performs correctly as sample size increases when
$\mathcal{H}_{0}$ is true. There is one small caveat, however, that needs to
be considered when $\mathcal{H}_{0}$ is false, namely, for large $n,$the model
will be identified as correct when $d_{CvM}(P_{true},F_{\theta_{0}})$%
\newline$<d_{i_{0}/M}.$ This underscores the need to identify what deviations
from $\mathcal{H}_{0}$ one wants to detect and then choosing $a$ so that
indeed such a failure can be detected.

\section{Examples}

In this section, the approach is illustrated through three examples, namely,
the location normal, location-scale normal, and the scale exponential models.
The effectiveness of the methodology is assessed using simulated samples from
a variety of distributions and in example 2 an application to a real data set
is presented. The following notation is used for the distributions in the
tables, namely, $N(\mu,\sigma^{2})$ is the normal distribution with mean $\mu$
and variance $\sigma^{2},t_{r}$ is the $t$ distribution with $r$ degrees of
freedom, exp$(\lambda)$ is the exponential distribution with mean $\lambda$
and $U(a,b)$ is the uniform distribution over $[a,b]$. For all cases we set
$r_{1}=r_{2}=1000$ in Algorithm B. We also provide the value $d_{min}%
=\inf_{\theta\in\Theta}d_{CvM}{\left(  F,F_{\theta}\right)  }$, where $F$ is
the true sampling distribution, as this indicates how close the true sampling
distribution is to the family $\left\{  F_{\theta}:\theta\in\Theta\right\}  $.
The R code \textquotedblleft distrMod" is used to calculate $d_{min}.$

For the simulations, samples of $n=20$ were generated from the distribution
$F$ in the table and then the methodology was applied to assess whether or not
the relevant model in the example is correct. Always the prior was taken to be
$DP\left(  a,F_{\theta(x)}\right)  $ except in Table \ref{tab2} where the
effect of making an inappropriate choice of $H$ is illustrated. Also, we
always took $p_{0}=1/M$ with $M=20$ so that $d_{p_{0}}=d_{0.05}$ is the
$0.05$-quantile of the prior distribution of $d_{CvM}.$ While one could always
choose $p_{0}$ smaller, the critical factor here is the choice of $a$ as the
prior has to be sufficiently concentrated about the family.$\smallskip$

\noindent\textbf{Example 1. }\textit{Location normal model.}

In this example $\{F_{\theta}:\theta\in\Theta\}=\{N(\theta,1):\theta
\in\mathcal{%
\mathbb{R}
\}}$ and so $\theta(x)=\bar{x}.$ In Table \ref{tab1} the relative belief
ratios and the strengths are recorded for testing the location normal model
against a variety of alternatives using several choices of the hyperparameter
$a.$ Recalling that we want $RB>1$ and the strength close to 1 when
$\mathcal{H}_{0}$ is true and $RB<1$ and the strength close to 0 when
$\mathcal{H}_{0}$ is false, it is seen that the methodology performs
wonderfully in every instance except one, namely, when the alternative is the
$t_{3}$ distribution and $a=1$. Surprisingly, the $t_{3}$ distribution has
distance from the location normal family equal to $d_{min}=0.0298$ which is
quite a bit smaller than the other alternatives. It is obviously more
difficult to detect model failure when this distance is small than otherwise. The solution to this, however, is seen from the table as this failure is detected for larger values of $a$. So to detect small
deviations it is necessary to use a prior that is more concentrated and this
can be assessed a priori. Notice that in all other cases the appropriate
conclusion is reached with $a=1$.

Figure 1 provides plots of the density of the prior distance and the posterior
distance for some cases. It follows, for instance, from Figure 1 that the
posterior density of the distance is more concentrated about 0 than the prior
density of the distance when the model is correct but not to the same degree otherwise.%

\begin{table}[tbp] \centering
\begin{tabular}
[c]{|c|c|c|c|c|}\hline
Distribution & $d_{min}$ & $a$ & $d_{0.05}$ & $RB$ (Strength)\\\hline
$N(0,1)$ & $0.0000$ & $1$ & $0.0192$ & $16.56\,(1.00)$\\
&  & $5$ & $0.0056$ & $5.72\,(0.71)$\\
&  & $10$ & $0.0035$ & $3.84\,(1.00)$\\\hline
$N(10,1)$ & $0.0000$ & $1$ & $0.0187$ & $16.44\,(1.00)$\\
&  & $5$ & $0.0064$ & $8.44\,(0.62)$\\
&  & $10$ & $0.0035$ & $4.72\,(1.00)$\\\hline
$N(0,4)$ & $0.1139$ & $1$ & $0.0200$ & $0.00\,(0.00)$\\
&  & $5$ & $0.0059$ & $0.00\,(0.00)$\\
&  & $10$ & $0.0028$ & $0.00\,(0.00)$\\\hline
$N(0,9)$ & $0.1657$ & $1$ & $0.0192$ & $0.00\,(0.00)$\\
&  & $5$ & $0.0053$ & $0.00\,(0.00)$\\
&  & $10$ & $0.0030$ & $0.00\,(0.00)$\\\hline
$0.5N(-2,1)+0.5N(2,1)$ & $0.1975$ & $1$ & $0.0184$ & $0.16\,(0.01)$\\
&  & $5$ & $0.0053$ & $0.00\,(0.00)$\\
&  & $10$ & $0.0034$ & $0.00\,(0.00)$\\\hline
$t_{0.5}$ & $0.5773$ & $1$ & $0.0208$ & $0.00\,(0.00)$\\
&  & $5$ & $0.0057$ & $0.00\,(0.00)$\\
&  & $10$ & $0.0031$ & $0.00\,(0.00)$\\\hline
$t_{3}$ & $0.0298$ & $1$ & $0.0192$ & $3.20\,(0.84)$\\
&  & $5$ & $0.0062$ & $0.48\,(0.11)$\\
&  & $10$ & $0.0030$ & $0.16\,(0.01)$\\\hline
$\text{Cauchy}(0,1)$ & $0.0755$ & $1$ & $0.0179$ & $0.00\,(0.00)$\\
&  & $5$ & $0.0057$ & $0.00\,(0.00)$\\
&  & $10$ & $0.0036$ & $0.00\,(0.00)$\\\hline
$U[0,1]$ & $0.1763$ & $1$ & $0.0192$ & $0.00\,(0.00)$\\
&  & $5$ & $0.0062$ & $0.00\,(0.00)$\\
&  & $10$ & $0.0029$ & $0.00\,(0.00)$\\\hline
$U[-1,1]$ & $0.0797$ & $1$ & $0.0208$ & $9.68\,(1.00)$\\
&  & $5$ & $0.0054$ & $0.16\,(0.02)$\\
&  & $10$ & $0.0014$ & $0.00\,(0.00)$\\\hline
$exp(1)$ & $0.0733$ & $1$ & $0.0200$ & $0.60\,(0.05)$\\
&  & $5$ & $0.0058$ & $0.04\,(0.00)$\\
&  & $10$ & $0.0029$ & $0.00\,(0.00)$\\\hline
$exp(10)$ & $0.2390$ & $1$ & $0.0179$ & $0.00\,(0.00)$\\
&  & $5$ & $0.0054$ & $0.00\,(0.00)$\\
&  & $10$ & $0.0032$ & $0.00\,(0.00)$\\\hline
\end{tabular}
\caption{Relative belief ratios and strengths for testing the location normal model with various alternatives and choices of
$a$ in Example 1.}\label{tab1}%
\end{table}%

\begin{figure}[h]
\centering
\subfigure [Example 1: $N(0,1)$] {\epsfig{figure=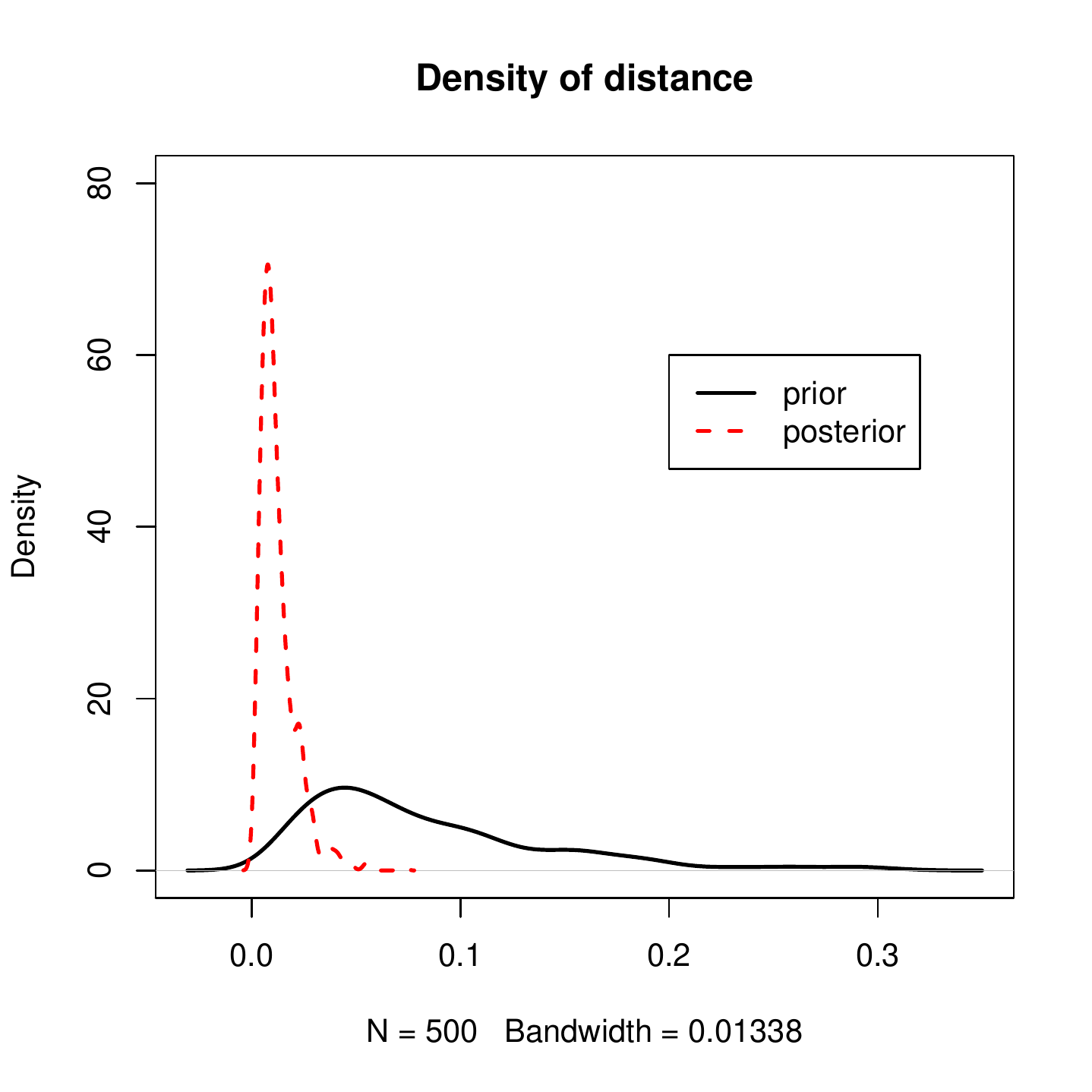,width=2.3in}}
\subfigure[Example 1: $t_{0.5}$]{\epsfig{figure=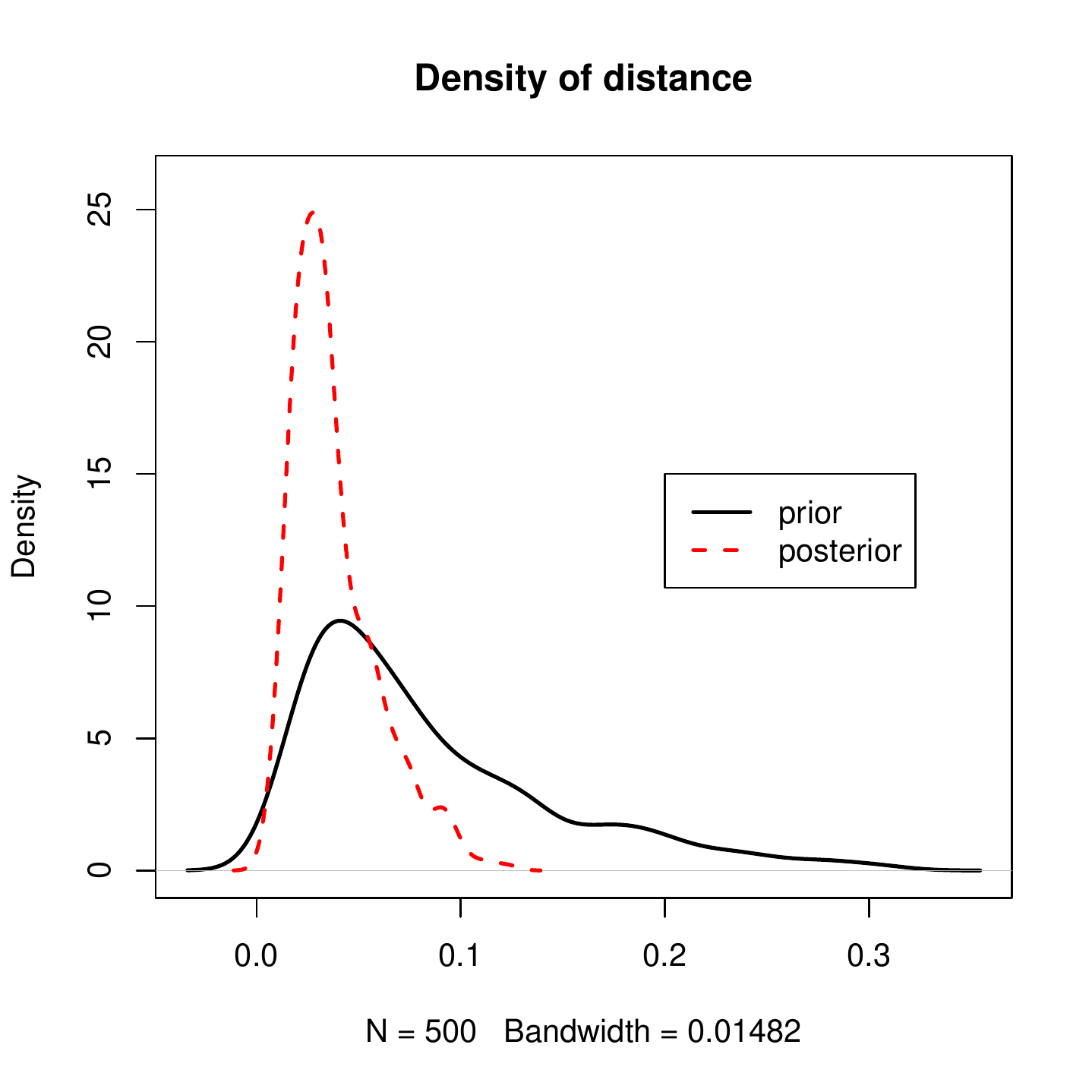,width=2.3in}}
\par
\subfigure [Example 1: $t_3$] {\epsfig{figure=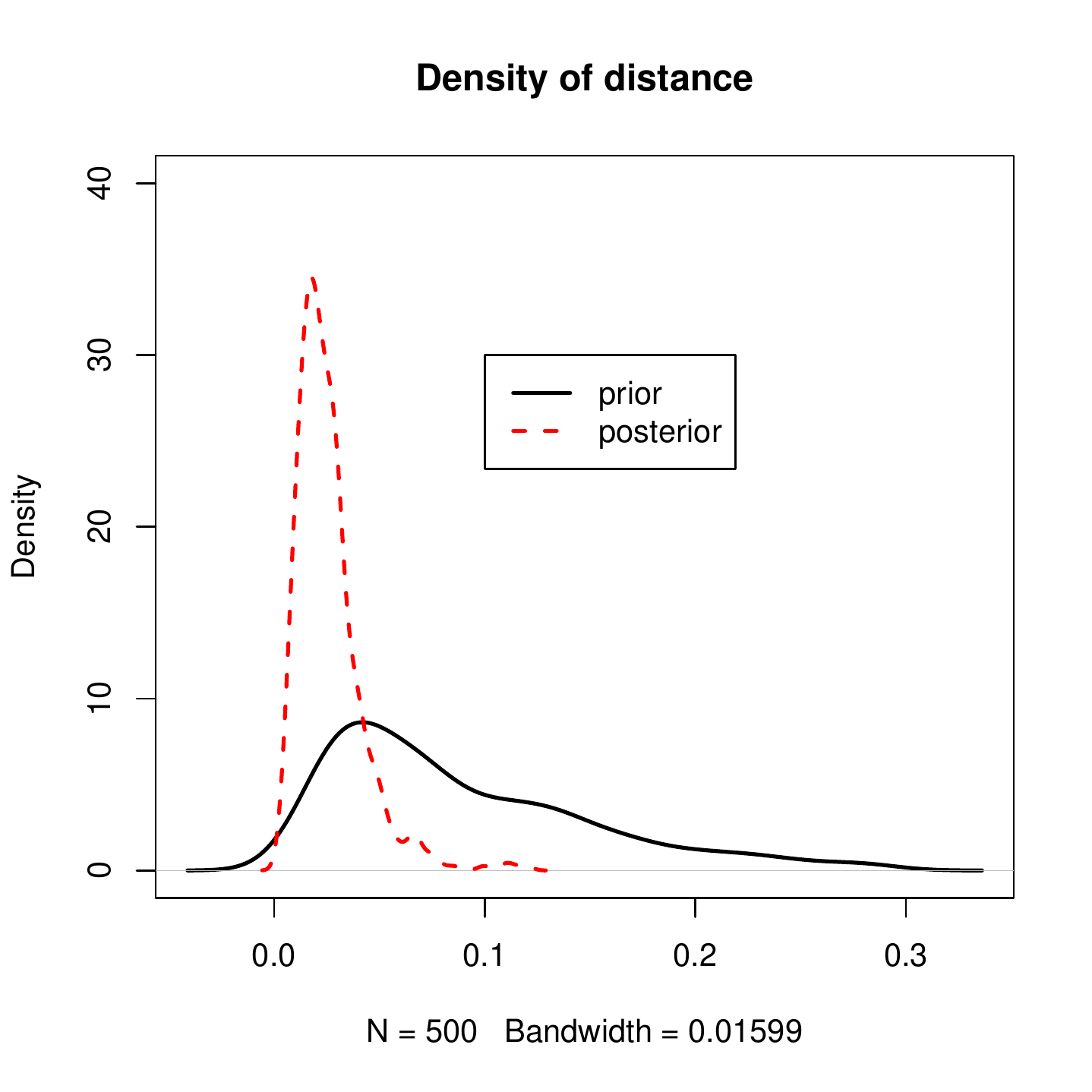,width=2.3in}}
\subfigure[Example 1: $\text{Cauchy}(0,1)$]{\epsfig{figure=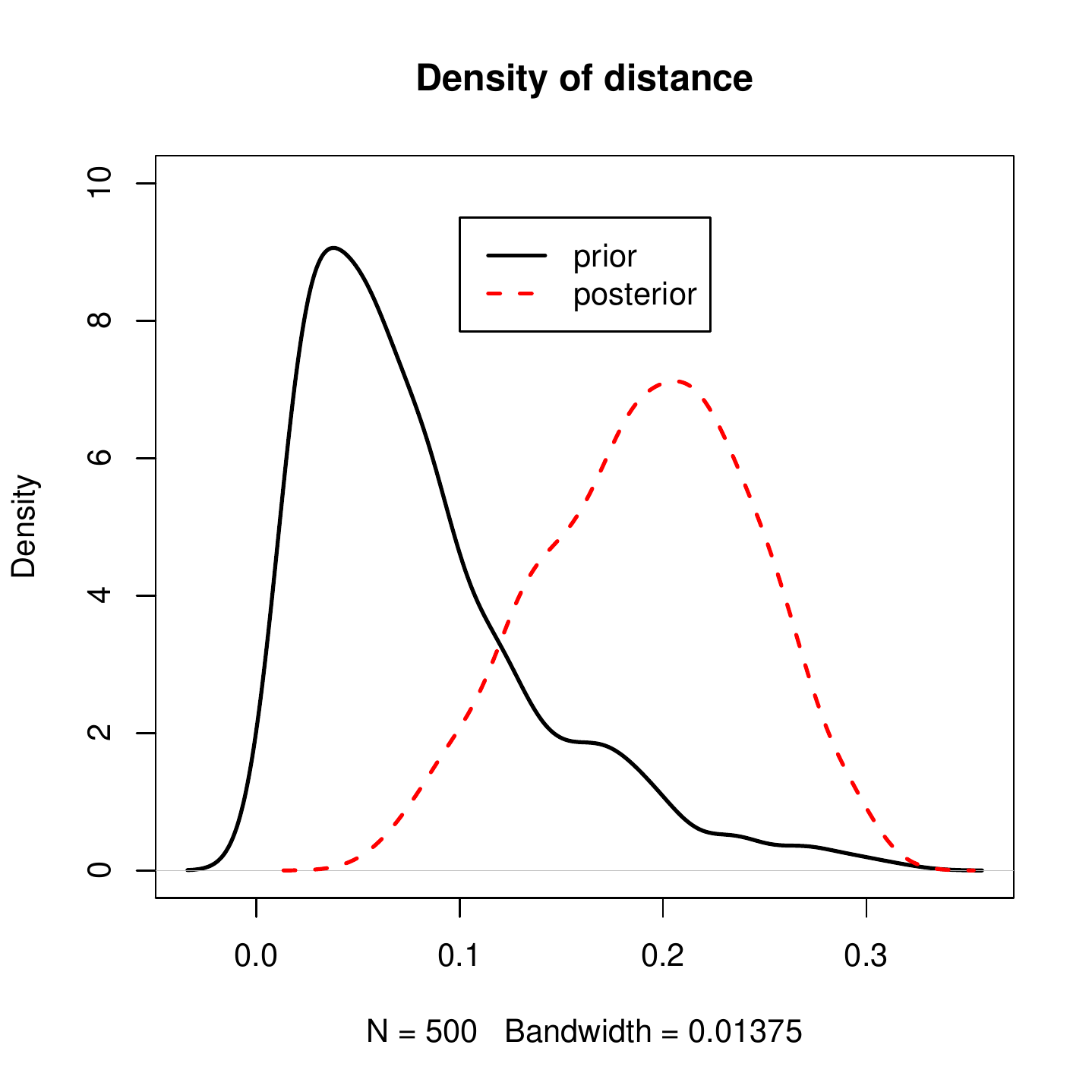,width=2.3in}}\caption{Plots of prior density versus posterior density of distance for some cases in Example 1.}%
\label{fig:SubF1}%
\end{figure}

It is interesting to consider the effect of prior-data conflict on the
methodology as this illustrates the importance of an appropriate choice of $H$
in the $DP(a,H)$ prior. Table \ref{tab2} gives the outcomes of model checking
for a particular sample of $n=20$ from the $N(10,1)$ distribution where
$\theta(x)=10.056$ was obtained and where various choices of $H$ and $a$ are
made. Clearly when $H=F_{\theta(x)}$, we get the correct conclusion about the
location normal model but not otherwise even though each $H$ is in the
location normal family. If $a$ is increased when $H$ is far from the truth,
this increases prior-data conflict and its ill effects.\smallskip%

\begin{table}[tbp] \centering
\begin{tabular}
[c]{|c|c|c|c|}\hline
Distribution & $a$ & $d_{0.05}$ & $RB$ (Strength)\\\hline
\multicolumn{1}{|c|}{$N(\theta(x),1)$} & $1$ & $0.0166$ & $17.20\,(1.00)$\\
\multicolumn{1}{|c|}{} & $5$ & $0.0053$ & $7.00\,(0.65)$\\
\multicolumn{1}{|c|}{} & $10$ & $0.0029$ & $3.20\,(0.84)$\\\hline
\multicolumn{1}{|c|}{$N(0,1)$} & $1$ & $0.0210$ & $0.00\,(0.00)$\\
\multicolumn{1}{|c|}{} & $5$ & $0.0053$ & $0.00\,(0.00)$\\
\multicolumn{1}{|c|}{} & $10$ & $0.0031$ & $0.00\,(0.00)$\\\hline
\multicolumn{1}{|c|}{$N(9,1)$} & $1$ & $0.0189$ & $0.04\,(0.00)$\\
\multicolumn{1}{|c|}{} & $5$ & $0.0056$ & $0.00\,(0.00)$\\
\multicolumn{1}{|c|}{} & $10$ & $0.0033$ & $0.00\,(0.00)$\\\hline
\end{tabular}
\caption{Relative belief ratios and strengths for testing the location normal model with various alternatives and choices of
$a$ in Example 1 when there is prior-data conflict.}\label{tab2}%
\end{table}%

\noindent\textbf{Example 2.} \textit{Location-scale normal model.}

In this example $\{F_{\theta}:\theta\in\Theta\}=\{N(\mu,\sigma^{2}%
):\theta=(\mu,\sigma^{2})\in\mathcal{%
\mathbb{R}
}\times(0,\infty)\}$ and so $\theta(x)=(\bar{x},\sum_{i=1}^{n}(x-\bar{x}%
)^{2}/n).$ The results are reported in Table \ref{tab3}. It is seen that in
all cases where the normal is correct the methodology gives the correct
answer. Failures occur with the mixture of normals and the $U[-1,1]$
distributions, as evidence is not obtained against the model in these cases.
In these cases the Cram\'{e}r-von Mises distance does not appear to give a
particularly powerful test against these alternatives. When the sample size
$n$ and $a$ are increased, however, model failure is detected. For example,
with $n=100$ and $a=20,$ the relevant relative belief ratios (strengths) are
$0.64\,(0.09)$ and $0.60\,(0.082)$ for the mixture of normals and $U[-1,1]$
distributions, respectively. So reasonably strong evidence is obtained against
normality in both cases and even more conclusive results are obtained with
$a=25,$ namely, $0.52(.06)$ and $0.16\,(0.02),$ respectively. %

\begin{table}[tbp] \centering
\begin{tabular}
[c]{|c|c|c|c|c|}\hline
Distribution & $d_{min}$ & $a$ & $d_{0.05}$ & $RB$ (Strength)\\\hline
$N(0,1)$ & $0.0000$ & $1$ & $0.0189$ & $17.52\,(1.00)$\\
&  & $5$ & $0.0061$ & $9.44\,(1.00)$\\
&  & $10$ & $0.0035$ & $3.84\,(1.00)$\\\hline
$N(10,1)$ & $0.0000$ & $1$ & $0.0172$ & $18.12\,(1.00)$\\
&  & $5$ & $0.0057$ & $10.52\,(1.00)$\\
&  & $10$ & $0.0035$ & $4.72\,(1.00)$\\\hline
$N(0,4)$ & $0.0000$ & $1$ & $0.0192$ & $15.40\,(1.00)$\\
&  & $5$ & $0.0064$ & $5.24\,(1.00)$\\
&  & $10$ & $0.0033$ & $1.92(0.792)$\\\hline
$N(0,9)$ & $0.0000$ & $1$ & $0.0184$ & $15.88\,(1.00)$\\
&  & $5$ & $0.0034$ & $3.00\,(0.382)$\\
&  & $10$ & $0.0030$ & $1.44\,(0.362)$\\\hline
$0.5N(-2,1)+0.5N(2,1)$ & $0.0462$ & $1$ & $0.0191$ & $16.52\,(1.00)$\\
&  & $5$ & $0.0055$ & $4.60\,(0.77)$\\
&  & $10$ & $0.0030$ & $2.28\,(0.75)$\\\hline
$t_{0.5}$ & $0.0575$ & $1$ & $0.0182$ & $0.00\,(0.00)$\\
&  & $5$ & $0.0057$ & $0.00\,(0.00)$\\
&  & $10$ & $0.0032$ & $0.00\,(0.00)$\\\hline
$t_{3}$ & $0.0120$ & $1$ & $0.0187$ & $13.84\,(0.308)$\\
&  & $5$ & $0.0062$ & $4.68\,(0.766)$\\
&  & $10$ & $0.0031$ & $0.96\,(0.356)$\\\hline
$\text{Cauchy}(0,1)$ & $0.0335$ & $1$ & $0.0198$ & $0.00\,(0.00)$\\
&  & $5$ & $0.00596$ & $0.00\,(0.00)$\\
&  & $10$ & $0.0031$ & $0.00\,(0.00)$\\\hline
$U[0,1]$ & $0.0246$ & $1$ & $0.0181$ & $17.68\,(1.00)$\\
&  & $5$ & $0.0060$ & $10.56\,(1.00)$\\
&  & $10$ & $0.0032$ & $1.00\,(0.00)$\\\hline
$U[-1,1]$ & $0.0245$ & $1$ & $0.0150$ & $17.6\,(0.12)$\\
&  & $5$ & $0.0051$ & $8.68\,(1.00)$\\
&  & $10$ & $0.0031$ & $6\,.00\,(0.70)$\\\hline
$exp(1)$ & $0.0567$ & $1$ & $0.0018$ & $0.12\,(0.01)$\\
&  & $5$ & $0.0053$ & $0.04\,(0.00)$\\
&  & $10$ & $0.0030$ & $0.00\,(0.00)$\\\hline
$exp(10)$ & $0.0567$ & $1$ & $0.0170$ & $0.20\,(0.00)$\\
&  & $5$ & $0.0055$ & $0.24\,(0.02)$\\
&  & $10$ & $0.0032$ & $0.20\,(0.01)$\\\hline
\end{tabular}
\caption{Relative belief ratios and strengths for testing the location-scale normal model with various alternatives and choices of
$a$ in Example 2.}\label{tab3}%
\end{table}%

Consider now the data of 100 stress-rupture lifetimes of Kevlar pressure
vessels presented in Andrews and Herzberg (1985). The goal is to test whether
the underlying distribution is normal. In this case $\theta
(x)=(209.171,37606.56)$. Previous studies such as Evans and Swartz (1994) and
Verdinelli and Wasserman (1998), suggested that model is not correct. In this
case $\inf_{\theta\in\Theta}d_{CvM}\left(  F_{n},F_{\theta}\right)  =$
$0.0494$, which is relatively a small distance, while $d_{CvM}\left(
F_{n},F_{\theta(x)}\right)  =0.0928$. The results in Table \ref{tab4} support
somewhat the non-normality of this data set although only when using a more
concentrated prior.  Figure 2 provides  plots  of the prior and posterior densities of the distance for various values of the concentration parameter $a$. It follows clearly from this figure that increasing the concentration parameter $a$  makes  the density of the prior distance more concentrated about $0$ than the density of the posterior distance. Thus, Figure 2 supports the conclusion of the non-normality of the data set.
\smallskip

\begin{table}[tbp] \centering
\begin{tabular}
[c]{|c|c|c|}\hline
$a$ & $d_{0.05}$ & $RB$ (Strength)\\\hline
$1$ & $0.0175$ & $18.68\,(1.00)$\\
$5$ & $0.0059$ & $4.96\,(1.00)$\\
$10$ & $0.0030$ & $0.56\,(0.02)$\\
$15$ & $0.0023$ & $0.28\,(0.01)$\\
$20$ & $0.0016$ & $0.08\,(0.01)$\\\hline
\end{tabular}
\caption{Relative belief ratios and strengths for testing the normality of the Kevlar data using various choices of
$a$ in Example 2.}\label{tab4}%
\end{table}%

\begin{figure}[h]
\centering
\subfigure [] {\epsfig{figure=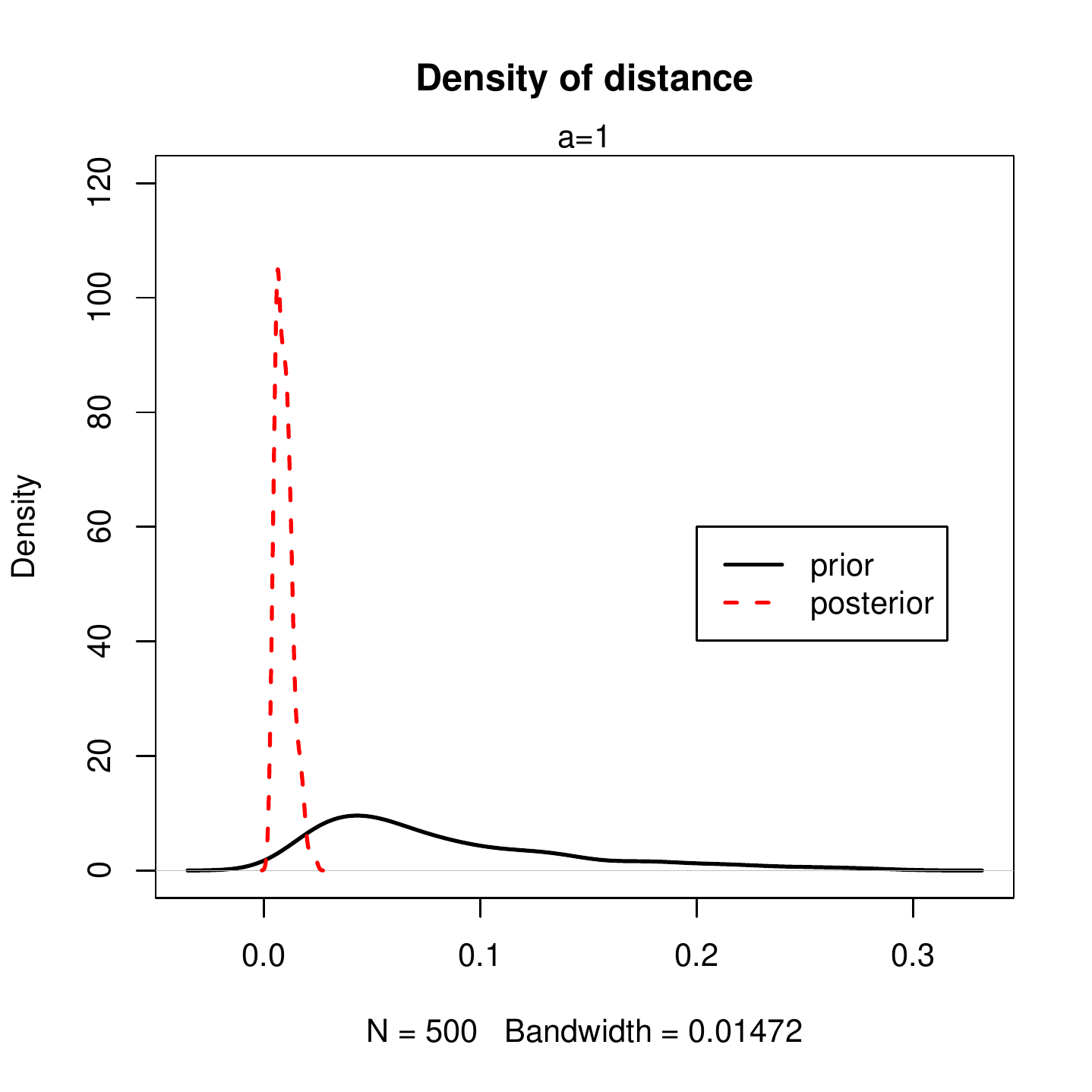,width=2.35in}}
\subfigure[]{\epsfig{figure=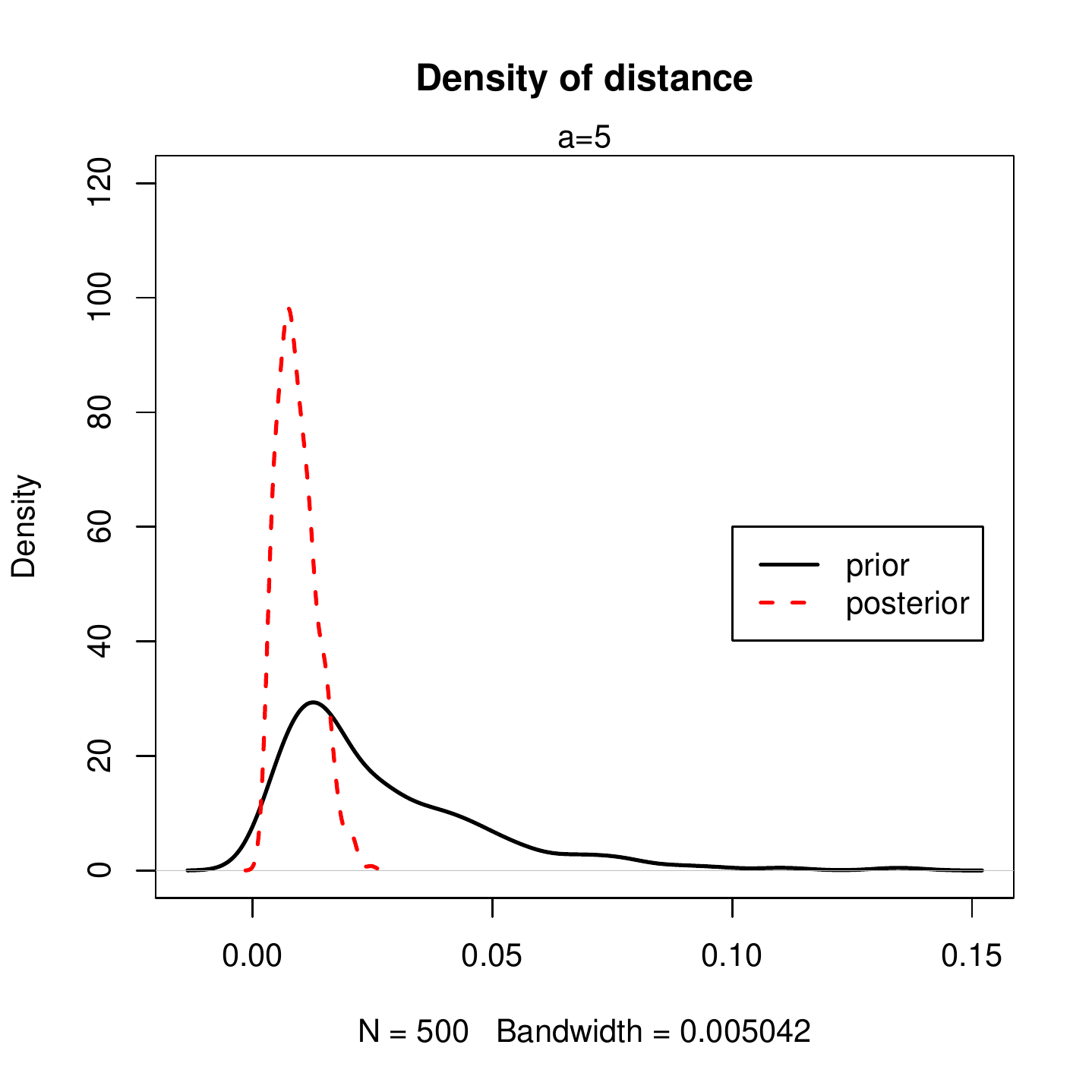,width=2.35in}}
\par

\subfigure [] {\epsfig{figure=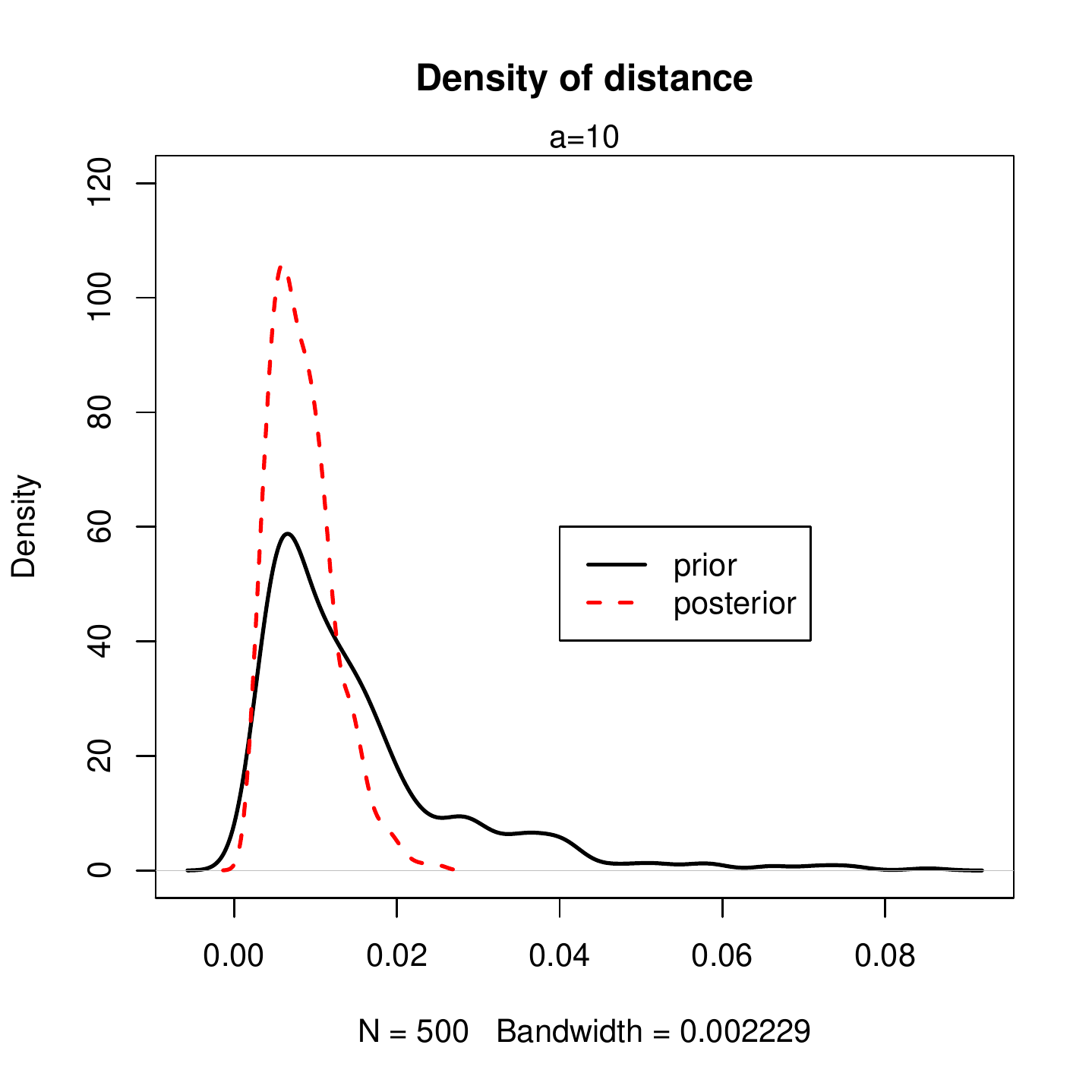,width=2.35in}}
\subfigure[]{\epsfig{figure=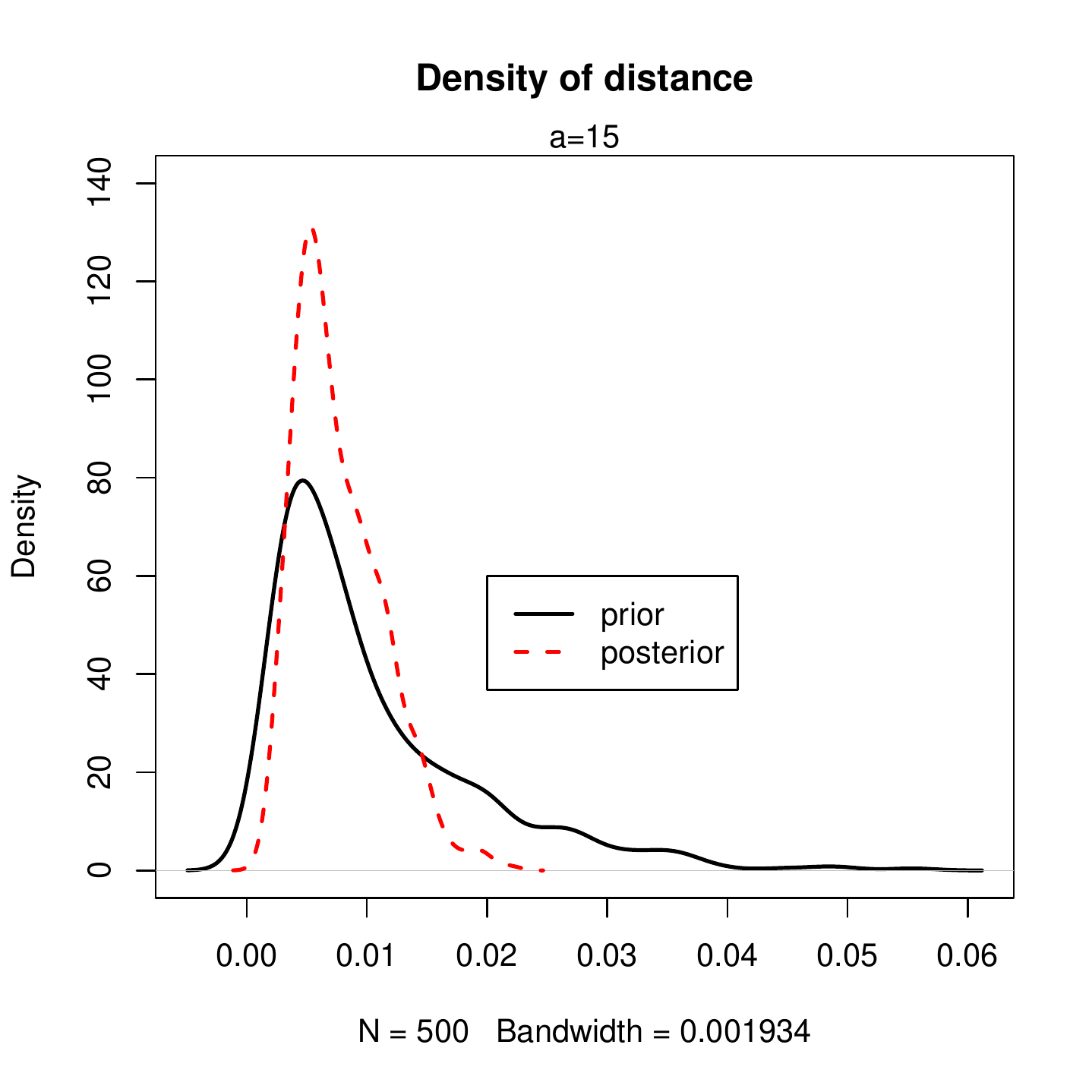,width=2.35in}}
\subfigure [] {\epsfig{figure=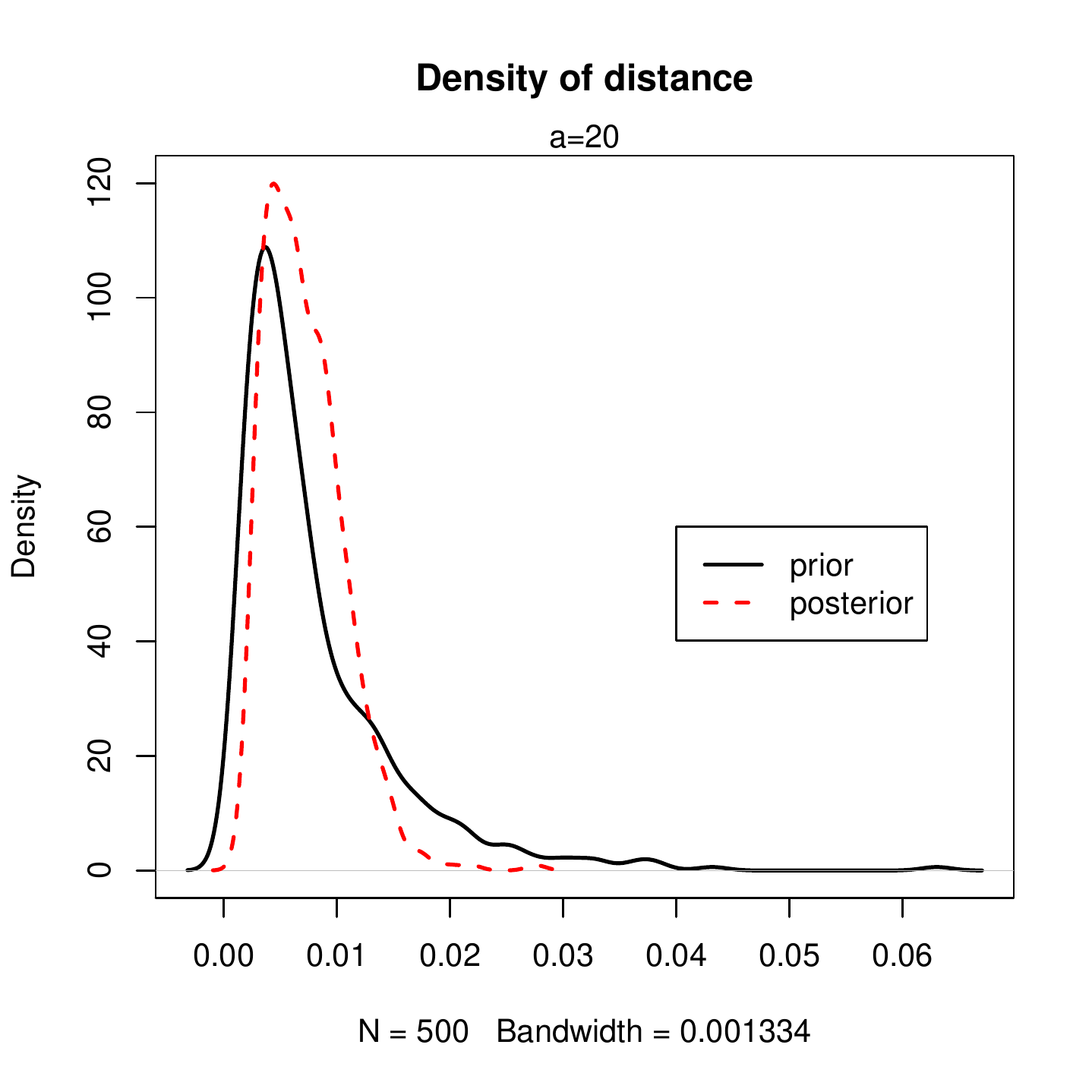,width=2.35in}}
\caption{(a) Plots of the prior and posterior densities of the distance for the stress-rupture lifetimes of Kevlar pressure
vessels data set discussed in Example 2  using various values of the concentration parameter $a$.}%
\label{fig:SubF1}%
\end{figure}

%

\noindent\textbf{Example 3.} \textit{Scale-exponential model.}

In this example $\{F_{\theta}:\theta\in\Theta\}=\{\exp(\theta):\theta
\in(0,\infty)\mathcal{\}}$ and so $\theta(x)=\bar{x}.$ The results are
reported in Table \ref{tab5} and it is seen that the methodology performs very
well here. In fact, the model is always correctly identified when it is true
and always strong evidence is obtained against the model when it is false
except when considering the $U[0,1]$ distribution with $a=1$ but the more
concentrated prior leads to evidence against.%

\begin{table}[tbp] \centering
\begin{tabular}
[c]{|c|c|c|c|c|}\hline
Distribution & $d_{min}$ & $a$ & $d_{0.05}$ & $RB$ (Strength)\\\hline
$N(0,1)$ & $0.2287$ & $1$ & $0.0176$ & $0.00\,(0.00)$\\
&  & $5$ & $0.0054$ & $0.00\,(0.00)$\\
&  & $10$ & $0.0030$ & $0.00\,(0.00)$\\\hline
$N(10,1)$ & $0.2780$ & $1$ & $0.0188$ & $0.00\,(0.00)$\\
&  & $5$ & $0.0056$ & $0.00\,(0.00)$\\
&  & $10$ & $0.0033$ & $0.00\,(0.00)$\\\hline
$N(0,4)$ & $0.2291$ & $1$ & $0.0189$ & $0.00\,(0.00)$\\
&  & $5$ & $0.0051$ & $0.00\,(0.00)$\\
&  & $10$ & $0.0029$ & $0.00\,(0.00)$\\\hline
$N(0,9)$ & $0.2292$ & $1$ & $0.01730$ & $0.00\,(0.00)$\\
&  & $5$ & $0.0063$ & $0.00\,(0.00)$\\
&  & $10$ & $0.0031$ & $0.00\,(0.00)$\\\hline
$0.5N(-2,1)+0.5N(2,1)$ & $0.2148$ & $1$ & $0.0179$ & $0.00\,(0.00)$\\
&  & $5$ & $0.0057$ & $0.00\,(0.00)$\\
&  & $10$ & $0.0033$ & $0.00\,(0.00)$\\\hline
$t_{0.5}$ & $0.2385$ & $1$ & $0.0187$ & $0.00\,(0.00)$\\
&  & $5$ & $0.0056$ & $0.00\,(0.00)$\\
&  & $10$ & $0.00284$ & $0.00\,(0.00)$\\\hline
$t_{3}$ & $0.2303$ & $1$ & $0.0186$ & $0.00\,(0.00)$\\
&  & $5$ & $0.0055$ & $0.00\,(0.00)$\\
&  & $10$ & $0.0031$ & $0.00\,(0.00)$\\\hline
$\text{Cauchy}(0,1)$ & $0.2336$ & $1$ & $0.0181$ & $0.00\,(0.00)$\\
&  & $5$ & $0.0053$ & $0.00\,(0.00)$\\
&  & $10$ & $0.0031$ & $0.00\,(0.00)$\\\hline
$U[0,1]$ & $0.0811   $ & $1$ & $0.0202$ & $3.60\,(1.00)$\\
&  & $5$ & $0.0060$ & $0.20\,(0.00)$\\
&  & $10$ & $0.0030$ & $0.08\,(0.01)$\\\hline
$U[-1,1]$ & $0.2266$ & $1$ & $0.0200$ & $0.00\,(0.00)$\\
&  & $5$ & $0.0060$ & $0.00\,(0.00)$\\
&  & $10$ & $0.0033$ & $0.00\,(0.00)$\\\hline
$exp(1)$ & $0$ & $1$ & $0.0173$ & $12.88\,(0.37)$\\
&  & $5$ & $0.0063$ & $4.76\,(1.00)$\\
&  & $10$ & $0.0032$ & $2\,(1.00)$\\\hline
$exp(10)$ & $0$ & $1$ & $0.0185$ & $16.88\,(1.00)$\\
&  & $5$ & $0.0057$ & $4.40\,(0.78)$\\
&  & $10$ & $0.0030$ & $1.16\,(0.30)$\\\hline
\end{tabular}
\caption{Relative belief ratios and strengths for testing the scale exponential model with various alternatives and choices of
$a$ in Example 3.}\label{tab5}%
\end{table}%

\section{Conclusions}

A general methodology for model checking based on the use of the Dirichlet
process and relative belief has been considered. This combination is seen to
lead to some unique advantages for this problem and this has been demonstrated
by developing theoretical properties of the procedure. Through several
examples, it has been shown that the approach performs extremely well.

While Cram\'{e}r-von Mises distance has been used here, other distance
measures could be used instead and may have distinct advantages in some
problems. For instance, the Anderson-Darling distance and the Kullback-Leibler
distance are possible substitutes. This entails simply substituting such
alternatives for $d_{CvM}$ in the algorithms. An important extension is the
generalization of the approach to construct tests for families of multivariate
distributions. While conceptually similar, there are computational and
inferential issues that need to be addressed and this is the subject of
current research.

\section{References}

\noindent Al-Labadi, L., and Zarepour, M. (2013). A Bayesian nonparametric
goodness of fit test for right censored data based on approximate samples from
the beta--Stacy process. \emph{Canadian Journal of Statistics}, 41, 3,
466--487. \smallskip

\noindent Al-Labadi, L., and Zarepour, M. (2014). Goodness of fit tests based
on the distance between the Dirichlet process and its base measure.
\emph{Journal of Nonparametric Statistics}, 26, 341-357.\smallskip

\noindent Andrews, D. F. and Herzberg, A. M. (1985) \emph{Data - A Collection
of Problems from Many Fields for the Student and Research Worker}.
Springer.\smallskip

\noindent Baskurt, Z. , and Evans, M. (2013). Hypothesis assessment and
Inequalities for Bayes factors and relative belief ratios. Bayesian Analysis,
8, 3, 569-590.\smallskip

\noindent Berger, J. O., and Guglielmi, A. (2001). Bayesian testing of a
parametric model versus nonparametric alternatives. \emph{Journal of the
American Statistical Association}, {96}, 174--184. \smallskip

\noindent Bondesson, L. (1982). On simulation from infinitely divisible
distributions. \emph{Advances in Applied Probability}, {14}, 885-869.
\smallskip

\noindent Carota, C., and Parmigiani, G. (1996). On Bayes factors for
nonparametric alternatives. In \emph{Bayesian Statistics 5} (J. M. Bernardo,
J. . Berger, A. P. Dawid, and A. F. M., eds.) Smith. Oxford University Press,
London. \smallskip

\noindent Choi, K. , and Bulgren, W. G. (1988). An estimation procedure for
mixtures of distributions. \emph{Journal of the Royal Statistical Society}, B,
30, 444--460.\smallskip

\noindent Dasgupta, A. (2008). \emph{Asymptotic Theory of Statistics and
Probability}. Springer, New York.\smallskip

\noindent Evans, M. (2015). \emph{Measuring Statistical Evidence Using
Relative Belief}. Monographs on Statistics and Applied Probability 144, CRC
Press, Taylor \& Francis Group.\smallskip

\noindent Evans, M. and Moshonov, H. (2006). Checking for prior-data conflict.
Bayesian Analysis, 1, 4, 893-914.\smallskip

\noindent Evans, M. and Swartz, T. (1994). Distribution theory and inference
for polynomial-normal densities. \emph{Communications in Statistics--Theory
and Methods}, 23, 1123--1148. \smallskip

\noindent Ferguson, T. S. (1973). A Bayesian analysis of some nonparametric
problems. \emph{Annals of Statistics}, {1}, 209-230.\smallskip

\noindent Florens, J. P., Richard, J. F., and Rolin, J. M. (1996). Bayesian
encompassing specification tests of a parametric model against a nonparametric
alternative. Technical Report 9608, Universits\'{e} Catholique de Louvain,
Institut de statistique. \smallskip

\noindent Gibbs, A., and Su, E. F. (2002). On choosing and Bounding
Probability metrics. \emph{International Statistical Review}, 70, 419-435.
\smallskip

\noindent Hsieh, P. (2011). A nonparametric assessment of model adequacy based
on Kullback--Leibler divergence. \emph{Statistics and Computing}, 23,
149--162. \smallskip

\noindent Ishwaran, H., and Zarepour, M. (2002). Exact and Approximate Sum
Representations for the Dirichlet Process. \emph{The Canadian Journal of
Statistics}, 30, 269-283.\smallskip

\noindent James, L. F. (2008). Large sample asymptotics for the two-parameter
Poisson-Dirichlet process. In \emph{Pushing the Limits of Contemporary
Statistics: Contributions in Honor of Jayanta K. Ghosh}, ed. B. Clarke and S.
Ghosal, Ohio: Institute of Mathematical Statistics, 187-199. \smallskip

\noindent Lavine, M. (1992). Some aspects of P\'{o}lya tree distributions for
statistical modelling. \emph{Annals of Statistics}, 20, 1222--1235. \smallskip

\noindent McVinish, R., Rousseau, J., and Mengersen, K. (2009). Bayesian
goodness of fit testing with mixtures of triangular distributions.
\emph{Scandivavian Journal of Statistics}, 36, 337--354. \smallskip

\noindent Sethuraman, J. (1994). A constructive definition of Dirichlet
priors. \emph{Statistica Sinica}, {4}, 639-650. \smallskip

\noindent Swartz, T. B. (1999). Nonparametric goodness--of--fit.
\emph{{Communications in Statistics: Theory and Methods}}, 28, 2821--2841.
\smallskip

\noindent Verdinelli, I., and Wasserman, L. (1998). Bayesian goodness-of-fit
testing using finite-dimensional exponential families. \emph{Annals of
Statistics}, 26, 1215--1241. \smallskip

\noindent Viele, K., (2000). Evaluating fit using Dirichlet processes.
Technical Report 384, University of Kentucky, Dept. of Statistics. \smallskip

\noindent Wolpert, R. L., and Ickstadt, K., (1998). Simulation of L\'{e}vy
random fields. In \emph{Practical Nonparametric and Semiparametric Bayesian
Statistics}, ed. D. Day, P. Muller, and D. Sinha, Springer, 227-242.
\smallskip

\noindent Zarepour, M., and Al-Labadi, L. (2012). On a rapid simulation of the
Dirichlet process. \emph{Statistics \& Probability Letters}, 82, 5, 916-924.

\end{document}